\documentclass[journal]{journal}
\usepackage[none]{hyphenat}
\usepackage{eurosym}
\usepackage{subfig}
\usepackage{xcolor}
\usepackage[draft]{todonotes}   
\usepackage{multirow}
\usepackage{eurosym}
\usepackage{graphicx}
\usepackage{subfig}
\usepackage{epstopdf}
\usepackage{caption}
\usepackage{booktabs}
\usepackage{amsmath}

\usepackage{tikz}
\usepackage{tkz-berge}
\usetikzlibrary{patterns,arrows,decorations.pathreplacing,automata}

\usepackage{pgfplots}
\pgfplotsset{compat=1.6}
\pgfplotsset{soldot/.style={color=blue,only marks,mark=*}} \pgfplotsset{holdot/.style={color=blue,fill=white,only marks,mark=*}}
\usepackage{balance}
\usepackage{amsmath}
\usepackage{amsfonts}
\usepackage{soul}
\usepackage{csvsimple}
\usepackage{longtable}
\usepackage{hyperref}

\tikzset{every state/.style={minimum size=45pt}}

\begin{document}
	
	%
	\title{e-Fair: Aggregation in e-Commerce for \\ Exploiting Economies of Scale}
	
	\author{Pierluigi Gallo,~
		Francesco Randazzo,~ 
		and~Ignazio Gallo
		\thanks{P. Gallo is with DEIM, Universit\`{a} di Palermo, viale delle Scienze ed. 9, 90128 Palermo, Italy, e-mail: pierluigi.gallo@unipa.it}
		\thanks{I. Gallo is with Universit\`{a} dell'Insubria, Via Ravasi 2, 21100 Varese, Italy, e-mail: ignazio.gallo@uninsubria.it}
	}
	
	\maketitle
	\thispagestyle{empty}
	
	\begin{abstract}
		\boldmath
		In recent years, many new and interesting models of successful online business have been developed, including competitive models such as auctions, where the product price tends to rise, and group-buying, where users cooperate obtaining a dynamic price that tends to go down.
		We propose the e-fair as a business model for social commerce, where both sellers and buyers are grouped to maximize benefits.
		e-Fairs extend the group-buying model aggregating demand and supply for price optimization as well as consolidating shipments and optimize withdrawals for guaranteeing additional savings.
		e-Fairs work upon multiple dimensions: time to aggregate buyers, their geographical distribution, price/quantity curves provided by sellers, and location of withdrawal points. 
		We provide an analytical model for time and spatial optimization and simulate realistic scenarios using both real purchase data from an Italian marketplace and simulated ones.
		Experimental results demonstrate the potentials offered by e-fairs and show benefits for all the involved actors.
	\end{abstract}
	
	\section{Introduction}
	\label{s:introduction}
	\IEEEPARstart{I}{n} recent years, the expansion of e-commerce services has led to the creation of new Internet-based business models, including auctions and group buying.
	
	Online auctions are becoming very popular both in business-to-business (B2B) and in consumer markets.  
	Unlike fixed prices mechanisms (FPMs), which dominated pricing strategies in last decades, online auctions introduce dynamic pricing mechanisms (DPMs) where buyers also dynamically influence the sale price. 
	English and Vickrey's auctions are the most popular on the Internet, but several different auction schemes exist, some have just appeared recently. 
	The most popular online auction site, eBay, adopts English auctions, but allows suppliers to set a limit to the highest price can be offered for the object. 
	The price decision policy can even be fully assigned to buyers, as it happens in Priceline.com. 
	First, buyers propose their own price for flight tickets and hotel rooms; then, sellers decide whether to accept such prices, based on demand and asset availability.
	
	Recently, also Group Buying (GB) business models appeared on the Internet: buyers cluster in groups to obtain discounts on purchasing products and services.
	In some cases, this works under the condition that a minimum number of requested items, otherwise potential buyers cannot finalize the purchase. 
	GB permits single buyers to get discounts that are normally available only to wholesalers.
	Products are displayed on the website during a time frame, typically called \emph{auction cycle}.
	As more buyers join the group, unit price drops down, according to price / quantity function, predetermined by sellers. 
	This function can be explicitly revealed to potential buyers or be withheld.
	
	The unit price decreases with the number of bought goods, independently if they come from an aggregation of several buyers, or if a single buyer requests multiple objects.
	The motto behind this price policy is 'the more you buy, the more you save'.
	
	An exemplary price definition, from staples.it regarding paper stacks, provides  a unit price of 4.69~\euro, between 1 and 9 elements. Buying from 10 to 29 stacks, the price is 4.19~\euro, from 30 to 59 it falls to 3.69~\euro, and finally, it remains 3.09~\euro, for higher demands.
	
	A similar pricing trajectory was found on MobShop.com, a pioneer group buying, during an auction cycle \cite{kauffman2001}. 
	Mobshop was one of the several group-buying sites appeared in the US and European markets in the early 2000s, together with Mercata, CoShopper, and LetsBuyIt.
	Despite the brilliant idea of grouping buyers, alone it was not enough to survive the market. 
	In facts, all these platforms closed their doors after a short time, for bankruptcy or insufficient gains. 
	A deep  analysis on the causes of these failures is out of the scope of this paper, however, a common factor was the incapability to capture great discounts and the limited range of product availability\footnote{Further details on these failures are available on online news of the epoch: http://www.wsj.com/articles/SB97951268061999104, http://www.cnet.com/news/group-buying-site-mercata-to-shut-its-doors/}.
	
	Concepts of volume-based discounts and grouping still survive and recently reappeared in different clothes.
	These two aspects reside in contemporary portals: the volume-based discounts is behind the price policy in staples.it, the demand aggregation lays behind  groupon.com.
	Groupon uses a fixed price mechanism and does not allow  flexibility in the choice of offered product.
	Items are displayed for 24~h as deals-of-the-day, with a high discount rate, even more than 50\%. 
	If the minimum number of buyers is reached by 24~h, they receive a redeemable coupon, and the price is debited to their credit cards.
	
	The key aspect, that is surprisingly understudied,  regards the availability of products, therefore we study aggregations for \emph{both} buyers and sellers and their impact on the final unit price.
	Aggregating one product supplied by multiple vendors simultaneously increases the quantity of goods available.
	This allows to consider bigger quantities, maximize buyers' savings, and optimize shipments.
	This last aspect is often overlooked and shipment aggregations contribute to obtaining remarkable savings for buyers.
	Recently,many solutions appeared on the scene for shipment aggregation.
	These services consolidate shipments even from multiple senders and permit to reduce shipment expenses.
	On the other hand, buyers can decide to receive their parcel, not at their home, but at some pick-up points, named also Points of Presence (POPs). 
	In our vision, these can be successfully exploited for consolidating shipments that are addressed to neighbor buyers.
	
	We focus on well-known issues of traditional group buying and propose a double-side aggregation that, at the best of our knowledge, is the first methodology which simultaneously takes into account buyers and sellers.
	Our first contribution is the definition of e-fairs, electronic fairs, as key opportunities to aggregate buyers \emph{and} sellers around desired products, providing an optimal trade-off among contrasting aspects.
	We modeled the aggregation of sellers and buyers as an optimization problem and described the evolution of e-fairs by the mean of a general architecture that uses finite state machines. Finally, e-fairs evolve over time depending on buyers arrivals, which are positively influenced by the e-fair itself.
	
	
	\section{Related Work}
	\label{s:related}
	In recent years, many business models for e-commerce, including auctions, group buying, and cooperation mechanism among sellers has been studied in different contexts.
	Separately, several works study the aggregation and shipment consolidation. 
	We present recent advances on such business models and aggregation, which are the main pillars on which e-fairs are based on.
	
	Recent business models require formation of groups of buyers.
	Long-term group formation mechanisms, based on trust relationships and credit-based group negotiation were studied in \cite{breban2001, breban2002a, breban2002b} and \cite{yuan2004}.
	In \cite{hyodo2003} buyers were optimally allocated in different GB websites that sell similar or equal products using a genetic algorithm.
	The authors in \cite{chen2009} defined a mechanism for GB that permit buyers to cooperate by sharing information, in order to coordinate their bidding.
	Authors of \cite{li2010} introduce the concept of Combinatorial Coalition Formation (CCF), permitting buyers to announce reserve prices for combinations of products. 
	Together with the sellers' price / quantity curves, these reserve prices influence the formation of groups of buyers.
	In \cite{he2006}, a distributed mechanism allows buyers to use two purchasing strategies together: a heuristic bundle search algorithm (bundle-based discounts), which allows buyers to purchase goods in different bundles, and a distributed coalition formation scheme (volume-based discounts), in order to optimize the total cost of goods. 
	In the buyers' coalition scheme proposed in \cite{matsuo2005}, each buyer places a sealed bid for all possible sets of items with reservation prices, and after a deadline, the mechanism allocates bundles of items using the VCG, the Vickrey–Clarke–Groves auction.
	Group buying was proposed also for categories of products \cite{yamamoto2001}, and buyers' web browsing history can be proficiently used to recommend products \cite{chen2012b}.
	The work in \cite{boongasame2014} takes into account issues related to locations of buyers and sellers and defines a novel buyer coalition scheme for forming a group of buyers in different locations.
	In the patent \cite{shoham2003}, it was proposed a complex system for automatically aggregate a group of customers and also provides a mechanism to promote competition among sellers.
	%
	In all the works listed above, the aggregation has been studied mainly from the buyer's side. All of these studies have focused on group-formation mechanisms as well as their economic and statistics implications.
	Unlike these works, the new e-fair mechanism we propose optimizes both sales and shipments, thanks to the dual-side aggregation.

	A strength of the framework we proposed concerns the optimization phase of goods purchasing and their distribution.
	In this context, several mathematical programming models, applied to distribution problems, can be found in the literature. 
	In order to demonstrate the economic advantage of decentralized distribution, the authors in \cite{yoshizaki1996} applied a transshipment model to the distribution of fuel in hundreds of big mills nationwide.
	In \cite{herer2001} it was considered a dynamic two-location transshipment problem with deterministic demands and the aim is to determine replenishment quantities and how much to transship each period in order to minimize the total costs over a finite planning horizon.
	The authors in \cite{dondo2009} use a mixed integer linear programming model for transportation of goods from factory to customers by using distribution centers for transshipments.
	%
	By extending the approaches used in these works, our search for the minimum price for shipping and sale has been modeled as an assignment and a transshipment problem.
	
	Package delivery is also called "the last mile problem", as it should ensure efficiency in terms of price, speed, and quality. 
	Traditionally, parcels are shipped directly to buyers' homes. 
	Recently, an increasing number of merchants and couriers are implementing new solutions for the last mile delivery, such as pick-up points where buyers withdraw the parcel (e.g. Access Point Lockers). 
	Pick-up points are usually convenient for those buyers who cannot receive packages at home and are also an economical delivery method because it reduces shipping costs for couriers. 
	In \cite{punakivi2003} it was demonstrated the cost advantage of pick-up points, which can be at unattended locations or at attended facilities such as shops, bars, railway stations, schools, etc. 
	In recent years, several attended and unattended schemes have been introduced in UK \cite{rowlands2006}.
	Empirical studies were made in \cite{esser2006, esser2007}, and despite pick-up points were adopted by few people, their cognition and usage increased significantly. 
	Novel pick-up systems continuously appear, where buyers retrieve goods purchased online from dedicated, secure locations~\cite{weltevreden2008}. 
	
	A similar approach to our pop aggregation system is the shipment \emph{consolidation}.
	Consolidation is a method of shipping whereby an agent (third-party logistics provider) combines individual parcels from various shippers into one shipment made to a destination agent, for the benefit of preferential rates. 
	On arrival, the consolidation is then de-consolidated by the destination agent into its original component parcels and made available to consignees.
	These activities take place at the urban consolidation centers (UCC), defined in \cite{browne2005} as logistics facilities situated in relatively close proximity to the served geographic areas: city centers, towns or  specific sites, from which consolidated deliveries are carried out within that area.
	UCC strategies could provide cost savings under several settings,  \cite{chen2012a} and great economies of scale, with high numbers of buyers and suppliers, contributes reducing the total shipping costs.
	Consolidation is integrable in our model as an intermediate layer between sellers and pops.
	This is complementary to our approach and different vendors can make consolidation to the same POP, through a third-party logistics provider.
	%
	Our transshipment model takes into account the shipment optimization problem from sellers to final customers, introducing all available pick-up points directly in the optimization model.
	
	\section{Unity is strength}
	\label{s:aggregation-model}
	In this section, first we present aggregation and its elementary constituents from different perspectives, then we discuss our strategies to aggregate buyers and sellers as well as shipments and withdrawals.
	
	\subsection{Aggregating buyers and sellers}
	\label{ss:aggregating-b-and-s}
	We propose to aggregate both sellers and buyers as in fairs. 
	Traditionally, fairs are organized as gatherings for selling goods, sometimes as exhibitions. 
	Like fairs, our e-fairs include timing concepts, promote social interaction, as well as aggregate buyers and sellers in a (virtual) space. 
	Furthermore, e-fairs provide entertainment and stimulate buyers' curiosity  about their own shopping goals and dynamic savings.
	
	Analogously to fairs, e-fairs aggregate supply coming from competing sellers in a reduced virtual area give by one single portal. 
	This increments the attraction mass for potential buyers, because they can look for a wide range of products and services
	as in a multi-vendor marketplace, but, with extra benefits due to economy of scale.
	e-Fairs hide the complex optimization procedures to buyers permitting them to have an easy purchase experience. 
	
	Unlike physical social communities for joint purchases, e-fairs aggregate demand that can be concurrently satisfied by multiple sellers. 
	Seller-side aggregation is both competitive and cooperative.
	On one hand, sellers compete to be selected by the e-fair system as the best provider for that specific demand and called to actually supply goods.
	
	For a specific volume of requested goods, there is an optimal assignment to sellers that minimize costs and deep changes in this assignment may happen when the total request changes.
	Total e-fair demand is fulfilled by several sellers and in the sense, they collaborate to the same goal. 
	However, no explicit inter-seller cooperation is expected among competitors.
	
	Buyers request one or more items to the e-fair and are attracted by price incentives both in sales and shipment. 
	Even in the case of flat price-quantity diagrams, when sellers do not apply economy of scale, e-fairs guarantee savings due to shipping aggregation.
	
	e-Fairs are fully defined by identifying a set of buyers, a set of sellers and their requirements in terms of buying and selling conditions.
	The set of buyers and the set of sellers are created in different ways: the first spontaneously aggregate according to their purchases wishes and timing, the latter are systematically aggregated by the e-fair system.
	
	Buyers are required to be patient about their waiting time from when they join the e-fair and when it ends and goods are shipped and specify their willingness to retrieve goods at indicated locations.
	Sellers are required to provide price/quantity curves for their products.
	
	\begin{figure*}[t]
		\centering
		\subfloat[]{\includegraphics[width=0.9\columnwidth]{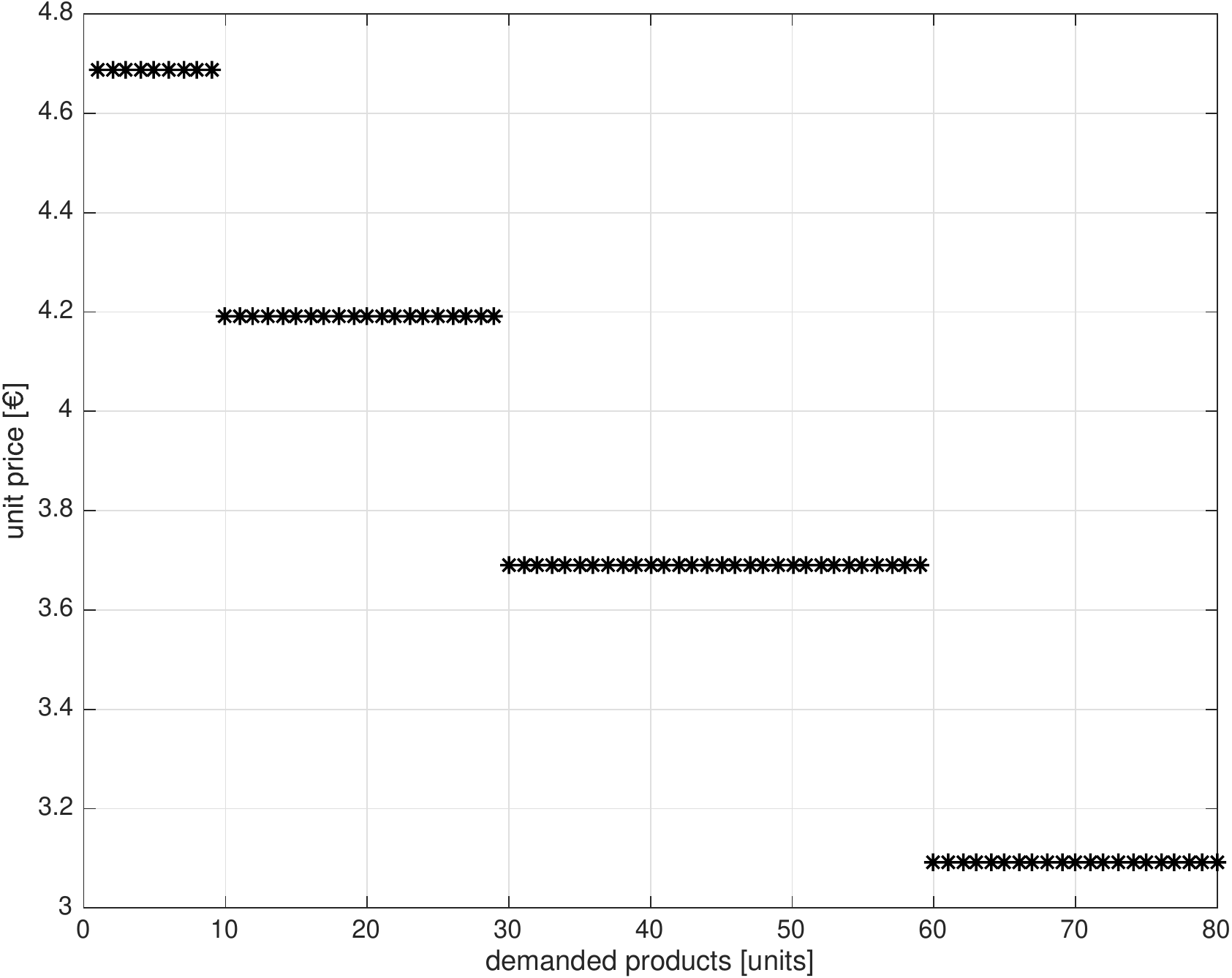}}\qquad
		\subfloat[]{\includegraphics[width=0.9\columnwidth]{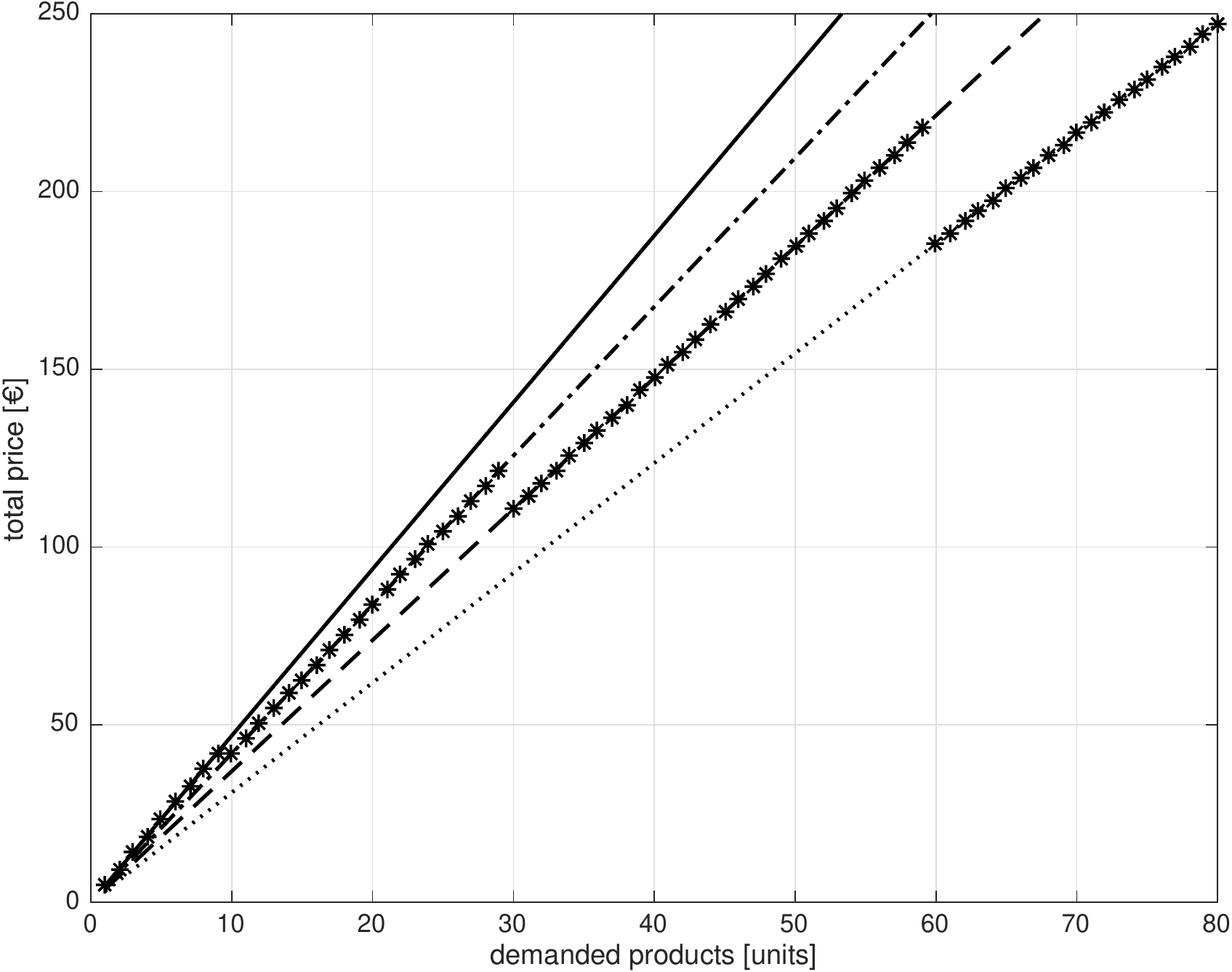}}
		\caption{Price diagrams with economy of scale. Unit price at Staples.it (a) and total price at Staples.it (b).}
		\label{f:fig_sim}
	\end{figure*}
	\subsection{Aggregating shipments and withdrawals}
	\label{sec:aggregating-shipments-and-withdrawals-using-points-of-presence}
	When e-fairs end, request and supply are unequivocally defined and shipment procedures start. 
	Shipment can be optimized by the means of aggregation, because it permits, again, an economy of scale.
	Instead of sending $n$ parcels to $n$ buyers at their location, we consider aggregating goods in a single parcel (eventually containing individually packaged goods). 
	This permits shipment savings for buyers because sellers usually reduce or even assume shipping costs when the ordered value is higher than a threshold. 
	On the other side, couriers save time and money because they manage only one delivery rather than several ones.
	
	Delivery aggregation requires also physical and geographic aggregation, i.e. the shipment has to be sent to one physical place.
	Sending goods to one of the buyers is infeasible because the extra responsibility given to the buyer chosen for shipment, which in general is not trusted.
	
	A possible solution, which is perfectly aligned with a recent trend, is shipping to a Point Of Presence (POP); buyers will be informed their goods are available and they may decide about withdrawal.
	Several commercial services have been made available in Italy to receive parcels on  behalf of buyers \cite{fermopoint,indabox,ioritiro,prontopacco}. 
	These service providers advertise the benefits in terms of time availability and security of the withdrawal, we foresee also the advantages of shipment aggregation. 
	
	\subsection{Cost components}
	The amount of money paid by a buyer for a purchase includes several cost components due to sales, shipment, and withdrawal.
	Sales costs depend on the number of products and on the seller, delivery costs are in general independent of distance but vary with the size of the parcel in volume and weight.
	Delivery transportation costs afforded by couriers to move goods from sellers to POPs and pick up costs afforded by buyers to withdraw goods from POPs are related to traveled distance, which has to be minimized.
	
	We indicate the e-fair as composed by a set of sellers $\{S_i | i=1,\dots ,I\}$, a set of POPs $\{P_j | j=1, \dots,J\}$ and a set of  buyers $\{B_k | k=1,\dots, K\}$. 
	Within the same country, the cost of delivery does not generally depend on the location of source and destination, we assume this cost as a parameter $S$, independent on the seller and the number of goods.
	
	Conversely, the pick up cost from position of $B_k$ to position of $P_j$, is proportional to the traveled distance $d(B_i,P_j)$.
	Therefore, $d_{jk}= \beta \cdot d(P_j,B_k)$, where $\beta$ is the average cost per kilometer incurred by the buyer (e.g. 0,15\euro / km).
	We assume that POPs and buyers are at fixed locations, therefore, reciprocal distances are given and unmodifiable values in an e-fair.
	
	\subsection{Economies of scale}
	Several companies already apply discounts related to volumes of sold products and services. 
	The general idea is that the higher the requested quantity, the lower is the unit price, being cost reductions due to an economy of scale.
	This general concept can be implemented in different ways through the most diverse price-quantity functions; an exemplary diagram is reported in Fig.~\ref{f:fig_sim} for one unit and total price for case paper sold by Staples.it. 
	
	Despite the simplicity of this unit price diagram, the trend of the total price presents surprising elements.  
	The diagram is defined as a piece-wise function whose segments have reduced slopes for higher numbers of units of products but, the total price diagram depicted in Fig.~\ref{f:fig_sim}b is neither continuous nor monotone.
	This introduces a weird behavior in proximity to discontinuities, where the total cost for a given number of goods is lower than the cost to buy more units. 
	For example, rather than buying a number of paper stacks between 50 and 59, it is better to order 60 units because they cost less.
	This counterintuitive behavior is not justified by the economy of scale and should be subject to attention from sellers.
	
	\section{The fair optimization model}
	\label{s:fair-based-model}
	
	\subsection{Modeling economies of scale}
	The economy of scale has a central role in our e-fair optimization model. 
	We use the discontinuous function for the total price described in previous section and move towards a continuous price-quantity curve, which is easier to manage for our optimization purposes.
	\begin{figure}
		\centering
		\includegraphics[width=0.9\columnwidth]{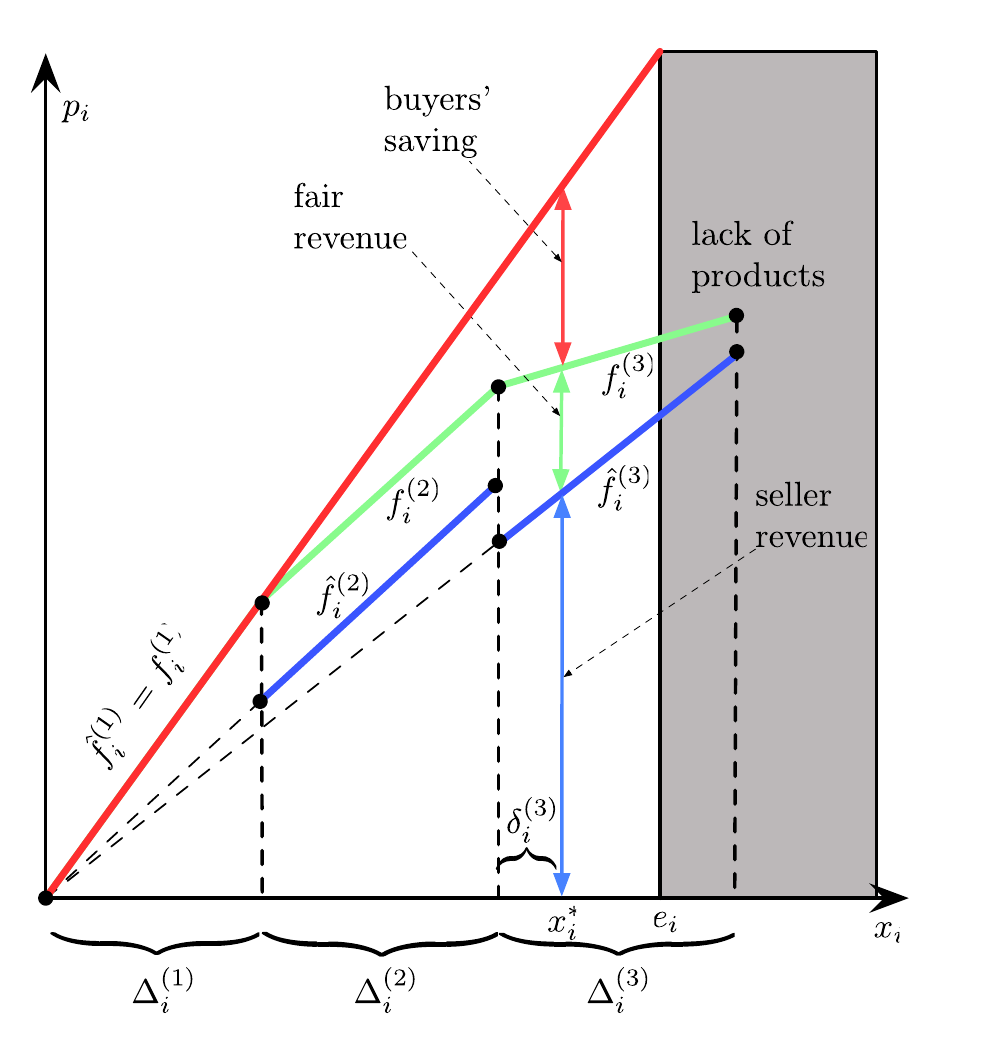}
		\caption{Total price functions for a number of products, considering the i-th seller: without e-fair (in red), with economy of scale (in blue) and  its continuous variant offered by the e-fair (in green).}
		\label{f:piece-wise}
	\end{figure}
	In Fig.~\ref{f:piece-wise} it is shown the total price. 
	Without e-fair the price grows linearly, as shown by the red line, then in blue, there is the discontinuous function for the total price that models the economy of scale from the seller's perspective. 
	The trend of total price managed by the e-fair is depicted in green, with a piece-wise linear continuous function.
	Blue segments are analogous to Fig.~\ref{f:fig_sim}b; these are provided by seller and their slopes are indicated with $ \hat{f}_i^{(l)} $.
	Slopes are the unit cost in the $l$-th range of quantities, therefore the total cost is $\hat{p}_i(x_i)=\hat{f}_i^{(l)} x_i, ~~~ \forall l \in {1,2,...,L}$, where $x_{i}$ is the total number of goods purchased at the end of a e-fair. 
	
	The green piecewise function can be expressed as follows:
	\begin{eqnarray}
		\label{eq:total-costs-fair}
		\begin{aligned}
			p_i(x_i)=
			\begin{cases} 
				f_i^{(1)} x_i     & \text{if }0 < x_i\leq x_i^{(1)} \\
				f_i^{(2)} x_i+c_2 & \text{if }x_i^{(1)} < x_i\leq x_i^{(2)} \\
				& \vdots \\
				f_i^{(L)} x_i+c_L & \text{if }x_i^{(L-1)} < x_i\leq x_i^{(L)} 
			\end{cases}
		\end{aligned}
	\end{eqnarray}
	where $f_i^{(l)}$ is the new unit cost of the $l$-th range and $c_l=\sum_{m=2}^l[(f^{(m-1)}-f^{(m)})x^{(m-1)}]$ for $l=2\dots L$ are opportunely added to remove discontinuities.
	The green diagram has to be lower limited by the blue diagram and upper limited by the red one.
	This is continuous and can be used withing a piece-wise linear optimization algorithm.
	
	Savings depend on $x^*_i$, the actual number of purchased products in the e-fair, as follows
	\begin{eqnarray}
		\label{eq:buyer_saving}
		s_i=f_i^{(1)} x^*_i - p_i(x^*_i)
	\end{eqnarray}
	The revenue for the e-fair manager is $p_i(x^*_i) - \hat{p}_i(x^*_i)$, marked  with the green arrow and the revenue for the seller is $\hat{f}_i^{(3)} x_i^*$, depicted with the blue arrow.
	The e-fair manager can tune values of $ f^{(l)}_{i} $ maintaining the green diagram limited by the red and blue ones, and having decreasing slopes moving to the right side of the figure.
	Tuning the green diagram is a trade-off: moving towards the blue one makes buyers' saving higher, moving it towards the red curve increases the fair manager gains.
	
	To express the total cost in a convenient analytical form that can be given to a mixed-integer linear optimization tool, we  introduce binary selection variables $w_i^{(l)} \in \{0,1\}$ and segment identifiers $\delta_i^l$ (with $l \in 1, \dots, L $).
	In this form, the original non-linear problem can be separated into linear functions, applying separable programming (see \S13.8 in \cite{HillLieb01}).
	$\delta^{(l)}_{i}$ is the number of goods that falls within the $l$-th interval. 
	The total number of objects requested to the seller $ S_i $ is
	\begin{eqnarray}
		\label{eq:num_requested_obj}
		x_{i} = \sum_{l=1}^L \delta^{(l)}_{i} 
	\end{eqnarray}
	and the total cost to be included in objective function for buying $ x_{i} $ goods becomes
	\begin{eqnarray}
		\label{eq:cost_requested_obj}
		\Phi_{i} = \sum_{l=1}^L f^{(l)}_{i}\delta^{(l)}_{i}
	\end{eqnarray}
	
	These variables have to meet the following constraints, considering $w_{i}$ as binary variables:
	\begin{eqnarray}
		\begin{aligned}
			\begin{cases} 
				\Delta^{(1)}_{i} w_{i}^{(1)} &\leq \delta^{(1)}_{i} \leq \Delta^{(1)}_{i} \\
				\Delta^{(2)}_{i} w_{i}^{(2)} &\leq \delta^{(2)}_{i} \leq \Delta^{(2)}_{i} w_{i}^{(1)} \\ 
				\Delta^{(3)}_{i} w_{i}^{(3)} &\leq \delta^{(3)}_{i} \leq \Delta^{(3)}_{i} w_{i}^{(2)} \\
				&\dots \\
				0 &\leq \delta^{(L)}_{i} \leq \Delta^{(L)}_{i}
			\end{cases}
		\end{aligned}
	\end{eqnarray}
	
	Defining $w^{(0)}_{i}=1$, the $I \times L$ constraints above can be generalized as follows, where $i=1,\cdots,I;~j=1,\cdots,L$:
	\begin{equation}
		\label{e:economy-of-scale-constraints-purchases}
		\Delta^{(l)}_{i} w_{i}^{(l)} \leq \delta^{(l)}_{i} \leq \Delta^{(l)}_{i} w_{i}^{(l-1)}
	\end{equation}
	
	From the set of inequalities in Eq.~\ref{e:economy-of-scale-constraints-purchases}, it results that   ${ w_{i}^{(l)} \leq  w_{i}^{(l-1)}}$, therefore when ${w_{i}^{(\underline{l})}=1}$, all ${w_{i}^{(l)} = 1}$ for ${l \leq \underline{l}}$ and ${w_{i}^{(l)} = 0}$ for ${l > \underline{l}}$. 
	This is graphically reported in Fig.~\ref{f:piece-wise}, where intervals are 'filled' from left to right and the rightmost partially filled interval has $w_i^{(\hat{l})}=0$.
	
	Constraint equations that model economy of scale for purchases apply also for shipments and withdrawals.
	Withdrawals systems generally use tickets.
	First, buyers have to purchase withdrawal tickets, then these can be spent when picking up goods.
	Analogously to Fig.~\ref{f:piece-wise} and Eq.~\ref{e:economy-of-scale-constraints-purchases}, 
	we indicate with $\gamma^{(m)}$ the number of tickets that falls within the m-th interval of the function price vs quantity of tickets and obtain the following expression (analogously to eq.~\ref{e:economy-of-scale-constraints-purchases}):
	\begin{equation}
		\label{e:economy-of-scale-constraints-withdrawal}
		\Gamma^{(m)} u^{(m)}\leq \gamma^{(m)} \leq \Gamma^{(m)} u^{(m-1)}.
	\end{equation}
	
	Unlike economy of scale on purchases, where $i$ indicates the i-th seller, economy of scale for buying tickets does not depend on POP and therefore has no subscript index.
	Conversely, this economy of scale depends on the withdrawal system in use (e.g. \cite{fermopoint,indabox}) because they can apply their own discount policy.
	However we omit such subscript for the sake of clarity and assume, without lack of generality, only one withdrawal system, being the multi-system case obtainable easily. 
	
	\begin{figure*}
		\centering
		\subfloat[]{\includegraphics[width=0.81\columnwidth]{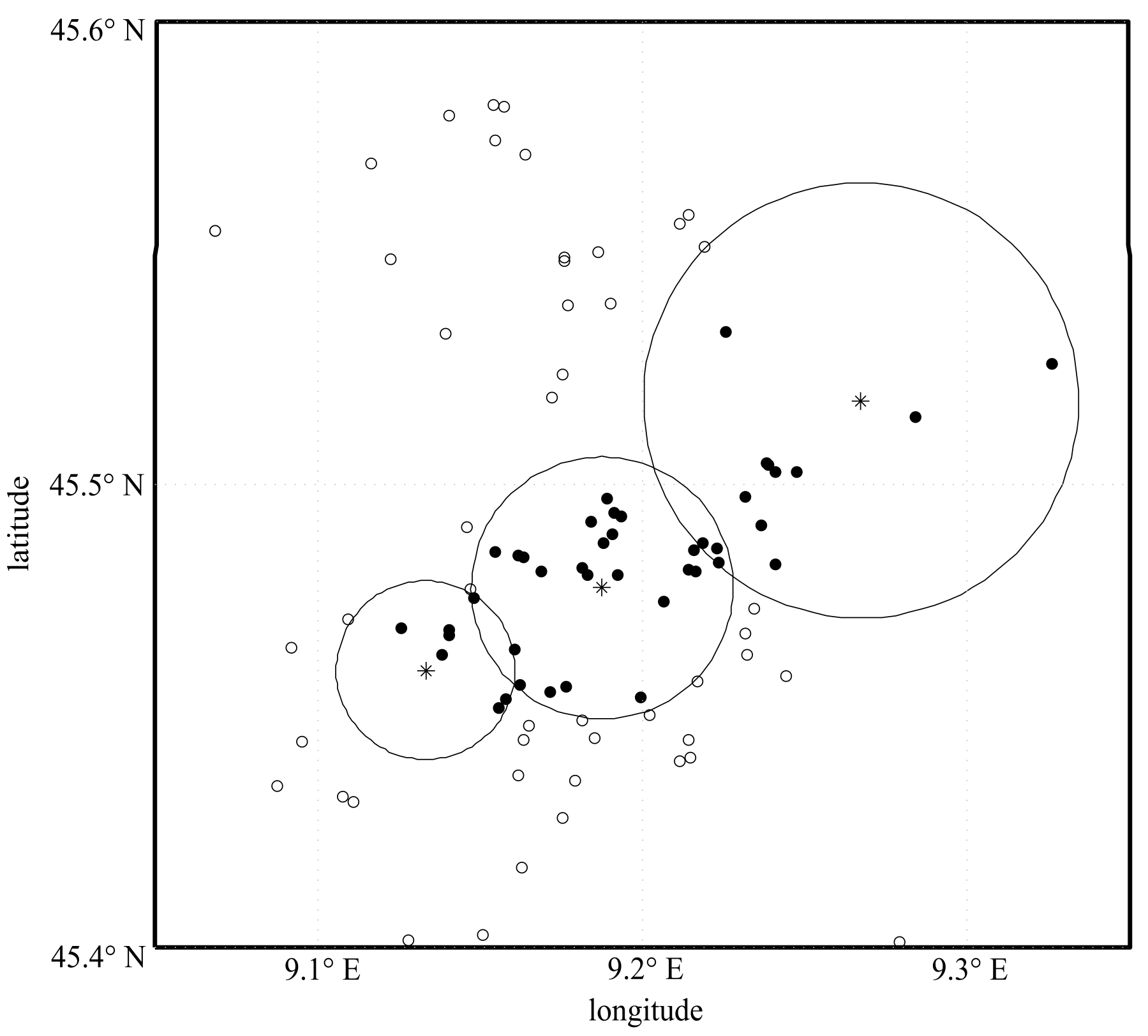}}\quad
		\subfloat[]{	\begin{tikzpicture}[scale=0.7,%
	VertexStyle/.style={shape=circle, draw,minimum size=12pt}]
	
	
	\begin{scope}[VertexStyle/.append style = {draw=none}]    
	\end{scope}
	
	\Vertex[L=$S_1$]{S1}
	\SO[unit=2,L=$S_2$](S1){S2}

	\begin{scope}[VertexStyle/.append style = {draw=none}] 
	\SO[unit=1.5,L=$\dots$](S2){S3} 
	\end{scope} 
		\EA[unit=3,L=$Fair$](S3){P2} 
	\SO[unit=1.5,L=$S_i$](S3){Si} 
	\begin{scope}[VertexStyle/.append style = {draw=none}]
	\SO[unit=1.5,L=$\dots$](Si){S4} 
	\end{scope} 
	\SO[unit=1.5,L=$S_I$](S4){SI}
	
	\tikzset{EdgeStyle/.style={->}} 
	\EdgeFromOneToSel[style=dashed]{S}{P}{1}{2} 
	\EdgeFromOneToSel[style=dashed]{S}{P}{2}{2} 
	\EdgeFromOneToSel[style=dashed]{S}{P}{I}{2} 
	\EdgeFromOneToSel[style=solid,lw=1.5pt,label=$\Phi_{1}$]{S}{P}{1}{2} 
	\EdgeFromOneToSel[style=solid,lw=1.5pt,label=$\Phi_{2}$]{S}{P}{2}{2}  
	\EdgeFromOneToSel[style=solid,lw=1.5pt,label=$\Phi_{2,2}$]{S}{P}{2}{2} 
	\EdgeFromOneToSel[style=solid,lw=1.5pt,label=$\Phi_{2}$]{S}{P}{2}{2}    
	\end{tikzpicture}}\quad 
		\subfloat[]{	\begin{tikzpicture}[scale=0.6,%
		VertexStyle/.style={shape=circle, draw,minimum size=12pt}]
		\Vertex[L=$B_1$]{B1}   
		\SO[unit=2,L=$B_2$](B1){B2}   
		\begin{scope}[VertexStyle/.append style = {draw=none}]
		\SO[unit=1,L=$\cdots$](B2){Bdots1}  
		\end{scope}
		\SO[unit=1,L=$B_k$](Bdots1){Bk} 
		\begin{scope}[VertexStyle/.append style = {draw=none}]    
		\SO[unit=1,L=$\dots$](Bk){Bdots2} 	
		\end{scope}
		\SO[unit=1,L=$B_K$](Bdots2){BK}   
		
		\WE[unit=5,L=$P_2$](B1){P2}   
		\NO[unit=2,L=$P_1$](P2){P1}  
		
		\begin{scope}[VertexStyle/.append style = {draw=none}]    
		\WE[unit=3.5,L=$x_{ij}$](P1){x} 	
		\EA[unit=3.5,L=$y_{jk}$](P1){y} 
		\end{scope}
		
		\SO[unit=2,L=$P_3$](P2){P3}
		\begin{scope}[VertexStyle/.append style = {draw=none}]    
		\SO[unit=1,L=$\dots$](P3){P5} 	
		\end{scope}
		\SO[unit=1,L=$P_j$](P5){Pj} 
		\begin{scope}[VertexStyle/.append style = {draw=none}]    
		\SO[unit=2,L=$\dots$](Pj){P6} 	
		\end{scope}
		
		\SO[unit=2,L=$P_J$](P6){PJ} 	
		\begin{scope}[VertexStyle/.append style = {draw=none}]    
		\WE[unit=3.5,L=$\alpha_{ij}$](PJ){a} 	
		\EA[unit=3.5,L=$\beta_{jk}$](PJ){b} 
		\end{scope}
		
		\WE[unit=5,L=$S_1$](P2){S1}   
		\SO[unit=2,L=$S_2$](S1){S2} 
		
		\begin{scope}[VertexStyle/.append style = {draw=none}] 
		\SO[unit=1,L=$\dots$](S2){S3} 
		\end{scope} 
		\SO[unit=1,L=$S_i$](S3){Si} 
		\begin{scope}[VertexStyle/.append style = {draw=none}]
		\SO[unit=1,L=$\dots$](Si){S4} 
		\end{scope} 
		\SO[unit=1,L=$S_I$](S4){SI}
		
		\tikzset{EdgeStyle/.style={->}} 
		\EdgeFromOneToSel[style=dashed]{S}{P}{1}{1,2,3,J} 
		\EdgeFromOneToSel[style=dashed]{S}{P}{2}{1,2,3,J} 
		\EdgeFromOneToSel[style=dashed]{S}{P}{I}{1,2,3,J} 
		
		
		\EdgeFromOneToSel[style=solid,lw=1.5pt,label=$\Phi_{1,1}$]{S}{P}{1}{1} 
		\EdgeFromOneToSel[style=solid,lw=1.5pt,label=$\Phi_{2,1}$]{S}{P}{2}{1}  
		\EdgeFromOneToSel[style=solid,lw=1.5pt,label=$\Phi_{2,2}$]{S}{P}{2}{2} 
		\EdgeFromOneToSel[style=solid,lw=1.5pt,label=$\Phi_{2,J}$]{S}{P}{2}{J}

		\EdgeFromOneToSel[style=dashed]{P}{B}{1}{1,2,K} 
		\EdgeFromOneToSel[style=dashed]{P}{B}{2}{1,2,K}
		\EdgeFromOneToSel[style=dashed]{P}{B}{3}{1,2,K}
		\EdgeFromOneToSel[style=dashed]{P}{B}{J}{1,2,K}
		\EdgeFromOneToSel[style=solid,lw=1.5pt,label=$\Psi_{1,1}$]{P}{B}{1}{1}  
		\EdgeFromOneToSel[style=solid,lw=1.5pt,label=$\Psi_{2,2}$]{P}{B}{2}{2}   
		\EdgeFromOneToSel[style=solid,lw=1.5pt,label=$\Psi_{2,K}$]{P}{B}{2}{K}  
		\end{tikzpicture}}\\
		\caption{
			first phase: filtering out unnecessary POPs due to distance ({\em pre-processing}) (a); 
			second phase: compute number $x_i$ of goods sold by each seller ({\em assignment problem}) (b);
			third phase: optimization based on a tripartite graph obtaining  $x_{ij}$ and $y_{jk}$  ({\em transhipment problem}) (c)}.
		\label{f:fair-model} 
	\end{figure*}

	\begin{figure*}
		\centering
		\scalebox{.6}{	\begin{tikzpicture}[node distance=2.5cm,->,>=latex,auto,
	every edge/.append style={thick}]
	\node[state] (0) {$0$};

	\node[state] (1) [right of=0] {$1$};
	\node[state] (2) [right of=1] {$2$}; 
	\node[state] (3) [right of=2] {$\cdots$}; 
	\node[state] (4) [right of=3] {$n_1$}; 
	\node[state] (5) [right of=4] {$n_1+1$}; 
	\node[state] (6) [right of=5] {$n_1+2$}; 
	\node[state] (7) [right of=6] {$\cdots$}; 
	\node[state] (8) [right of=7] {$n_2$}; 
	\node[state] (9) [right of=8] {$n_2+1$}; 
	\node[state] (10) [right of=9] {$\cdots$}; 
	\node[state] (11) [right of=10] {$n_L$}; 
	\path 
	(0) edge[bend left]  node{$\lambda_1$}   (1)
	(1) edge[bend left]  node{$\lambda_1$}   (2)
	(2) edge[bend left]  node{$\lambda_1$}   (3)
	(3) edge[bend left]  node{$\lambda_1$}   (4)
	
	(4) edge[bend left]  node{$\lambda_2$}   (5)
	(5) edge[bend left]  node{$\lambda_2$}   (6)
	(6) edge[bend left]  node{$\lambda_2$}   (7)
	(7) edge[bend left]  node{$\lambda_2$}   (8)
	(8) edge[bend left]  node{$\lambda_3$}   (9)
	(9) edge[bend left]  node{$\lambda_3$}   (10)
	(10) edge[bend left]  node{$\lambda_L$}   (11);

\draw [decorate,decoration={brace,amplitude=20pt},xshift=-4pt,yshift=0pt](0,1) -- (10,1)node [black,midway,xshift=9pt] {};

\draw [decorate,decoration={brace,amplitude=20pt},xshift=-4pt,yshift=0pt](10.5,1) -- (20,1)node [black,midway,xshift=9pt] {};

\draw [decorate,decoration={brace,amplitude=20pt},xshift=-4pt,yshift=0pt](20.5,1) -- (25,1)node [black,midway,xshift=9pt] {};

\draw [decorate,decoration={brace,amplitude=20pt},xshift=-4pt,yshift=0pt](25.5,1) -- (28,1)node [black,midway,xshift=9pt] {};
\end{tikzpicture}}
		\caption{Markov chain model for buyers' arrival in one e-fair}.
		\label{f:fair-model} 
	\end{figure*}
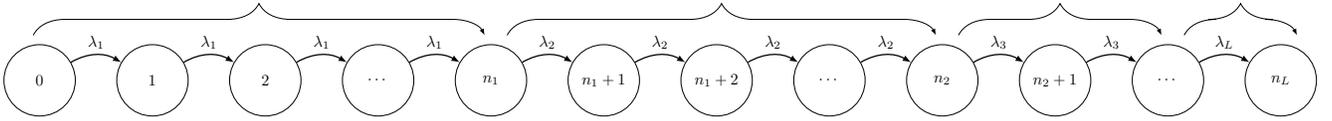
	
	We model fairs as an aggregation of sellers and buyers, saving purchase costs thanks to economy of scale, consolidating shipping and aggregating withdrawal procedures. 
	In our model, we consider $I$ sellers, $J$ pick-up points, and $K$ buyers.
	Sellers send goods to POPs, where buyers retrieve them (see Fig.~\ref{f:fair-model}c).
	We consider also the case when sellers ship directly to buyers' address yet maintaining the tripartite graph approach.
	We add K dummy POPs, whose position is buyers' home address. 
	In such a case, the withdrawal cost is zero because the good is shipped directly to the buyer. 
	For the sake of simplicity, such dummy POPs are not drawn in Fig.~\ref{f:fair-model}c.
	Goods move from sellers to buyers, along arches, tagged with shipped quantity $x_{ij}$ from i-th seller to j-th POP, and $y_{jk}$ products from j-th POP to the k-th buyer.
	Our goal is the minimization of the total cost, by defining an optimization problem that, given input data, decides optimal $x_{ij}$ and $y_{jk}$.
	This means taking decisions on the number of products to be shipped, the involved sellers, and locations of POPs.
	We defined a three-steps algorithm: after a preprocessing phase, purchase and logistics optimization are performed.
	
	\subsection{Optimization algorithm}
	\subsubsection{Pre-processing}
	Buyers that join an e-fair provide one shipment address (e.g. home, office, etc.) and additionally the maximum range they wish to retrieve goods, indicated as $d_{k,max}$.
	This maximum range is used in the preprocessing phase, where POPs that are out of reach for buyers are filtered out. 
	This happens for POPs whose distance $d(P_j,B_k) > d_{k,max}, \forall k$. 
	
	An exemplary prefiltering is shown in Fig.~\ref{f:fair-model}a for the metropolitan area around Milan.
	Buyers' shipment addresses are marked with asterisks. 
	We build around them circles with the radius equal to the maximum travel distance indicated by each buyer, therefore these circles have radius $d_{k,max}$. 
	Dots indicate POP positions, the white ones have been filtered out by preprocessing, and the remaining black ones can be used to aggregate shipment.
	Aggregation is possible for black dots that fall withing the intersection of more than one circle.
	
	Preprocessing reduces the number of POPs that are out of reach for buyers and their incident edges.
	This reduces the complexity of graph indicated in Fig.~\ref{f:fair-model}c.
	
	\subsubsection{Purchase optimization}
	In the second phase, we compute $x_{i}$ by minimizing the sum of purchase costs $\min_{\delta{i}}  \left\{  \sum_{i=1}^I \Phi_{i} \right\} $ (see Eq.~\ref{eq:cost_requested_obj}).
	This step considers the whole order from a specific seller and neglects details about the way this $x_i$ products have to be divided among POPs ($x_{ij}$ are computed during the next step). 
	This is known, in the Operation Research (OR) field, as \emph{Assignment problem} and it is shown in Fig.~\ref{f:fair-model}b,
	where the i-th seller contributes to the fair with $\Phi_i$ goods.
	
	\subsubsection{Shipment and withdrawal optimization}
	Finally, starting from $x_i$ obtained with step 2, we optimize shipment and obtain $x_{ij}$ and $y_{jk}$.
	This defines the distribution of goods among POPs and the retrieval plan for buyers. 
	This step works on the graphical model reported in Fig.~\ref{f:fair-model}c and is a \emph{Transshipment problem} in the OR field.  
	\begin{table*}[!t]
		\centering
		\caption{Constraints applied to the optimization model distinguished by second and third phases (upper and lower parts).}
		\label{tab:constraints}
		\begin{tabular}{r|c|l}
			\toprule
			Constraint expression & scope & Description \\
			\midrule
			\(\displaystyle \sum_{i=1}^{I} x_{i} =  y_{fair} 		\) & $\forall Fair$ 
			& The number of purchased products is equal to the number of products requested in the fair.\\
			\(\displaystyle \sum_{l=1}^{L} \delta_{i}^{(l)}  = x_i \)  & $\forall S_i$ 
			& The sum of variables of all the $L$ intervals is equal to the quantity of sold products by seller $S_i$. \\
			\(\displaystyle x_{i} \leq e_{i} 		\) & $\forall i$ 
			& The number of products sold  by seller $S_i$ is less than or equal his supply available.\\ \newline
			
			\(\displaystyle \Delta^{(l)}_{i} w_{i}^{(l)} \leq \delta^{(l)}_{i} \leq \Delta^{(l)}_{i} w_{i}^{(l-1)} \) & $\forall i,l$ 
			& Set of L inequalities which define the bounds of the $\delta_{i}^{(l)}$ quantities throw binary control variables $w_i$.\\
			\midrule
			\(\displaystyle \sum_{j=1}^{J} x_{ij} = x_{i}     \) & $\forall i$ 
			& The sum of shipped products in each trajectory is equal to the quantity of sold products by the seller $S_i$.\\
			
			\(\displaystyle x_{ij} 	\leq a_{ij} \cdot y_{fair}  \) & $\forall i,j$ 
			& When the arc $i \rightarrow j$ is used (i.e. $x_{ij} > 0$), the amount of products to be shipped from \emph{i}-th seller can be \\ & & 
			no larger than the total fair demand, which is always true. Consequently, shipment cost $S$ can be applied \\ & & only if the arc is used, because necessarily $a_{ij} = 1$.\\ 
			\(\displaystyle \sum_{i=1}^{I} x_{ij} \leq c_{j} 	\) & $\forall j$ 
			& The sum of products shipped to a POP is less than or equal to the POP capacity. \\
			\(\displaystyle \sum_{j=1}^{J} y_{jk} = y_{k}     	\) & $\forall k$ 
			& The quantity of products requested by the buyer is divided among POPs. \\
			\(\displaystyle \sum_{j=1}^{J} b_{jk} 	= 1  		\) &$ \forall k$ 
			& The buyer withdraws his products by only one POP.\\
			\(\displaystyle \sum_{i=1}^{I} x_{ij} 	= \sum_{k=1}^{K} y_{jk}  \)  & $\forall j $ 
			& The amount of incoming products at POP is equal to the amount of outgoing products (flow equation).\\
			
			\(\displaystyle \Gamma^{(m)} u^{(m)}\leq \gamma^{(m)} \leq \Gamma^{(m)} u^{(m-1)} \) & $ \forall m $
			& Set of M inequalities which define the bounds of the $\gamma^{(l)}$ quantities throw binary control variables $u$.\\
			\(\displaystyle y_{jk} \leq b_{jk} \cdot y_{fair}  \) & $\forall j,k$ 
			& When the arc $j \rightarrow k$ is used (i.e. $y_{jk} > 0$), the amount of products to be retrieved from \emph{k}-th buyer is no \\ & & larger than the total fair demand, which is always true. Consequently, pick up cost $d_{jk}$ can be applied \\ & & only if the arc is used, because necessarily $b_{jk} = 1$.\\ 
			\(\displaystyle \sum_{m=1}^M \gamma^{(m)} = \sum_{j=1}^J\sum_{k=1}^{K} b_{jk}  \) & $\forall j$ 
			& The sum of variables of all the $M$ intervals for the considered POP provider is equal to the total number \\ & & of withdrawals to POPs. \\
			\(\displaystyle x_{ij} \geq 0 						\) & $\forall i,j$ 
			& The quantity of products along each arc $i \rightarrow j$ is non-negative. \\
			\(\displaystyle y_{jk} \geq 0 						\) & $\forall j,k$ 
			& The quantity of products along each arc $j \rightarrow k$ is non-negative.\\
			\bottomrule
		\end{tabular}
	\end{table*}
	
	The objective function to be minimized contains several cost components:
	\begin{equation}
		\min_{a_{ij}, b_{jk}, x_{ij}, y_{jk}}  \left\{  \sum_{i=1}^I \sum_{j=1}^J \Phi_{ij} + \sum_{j=1}^J \sum_{k=1}^K \Psi_{jk} \right\} 
		\label{eq:total_fair_cost}
	\end{equation}
	where
	${\Phi_{ij} = \frac{\Phi_{i}}{x_{i}} x_{ij} +  S  a_{ij}}$ components are due to purchases and shipment, whereas ${\Psi_{jk} =\frac{y_{jk}}{\sum_k y_{jk}}  \sum_{m=1}^M g^{(m)} \gamma^{(m)}  + d_{jk}  b_{jk}}$ are given by withdrawals.
	Binary variables ${a_{ij},b_{jk} \in \{0,1\}}$ indicate, respectively if the $i\rightarrow j$ and $j\rightarrow k$ arches are used.
	
	This optimization model includes several equality and inequality constraints, reported in Table~\ref{tab:constraints}.
	The horizontal line separates constraints used in the second step from those used in the third step of the algorithm.
	
	\subsection{Buyers' time analysis}
	\label{buyers-arrival}
	\begin{figure*}[h!]
		\centering
		\subfloat[]{\includegraphics[height=7.92cm]{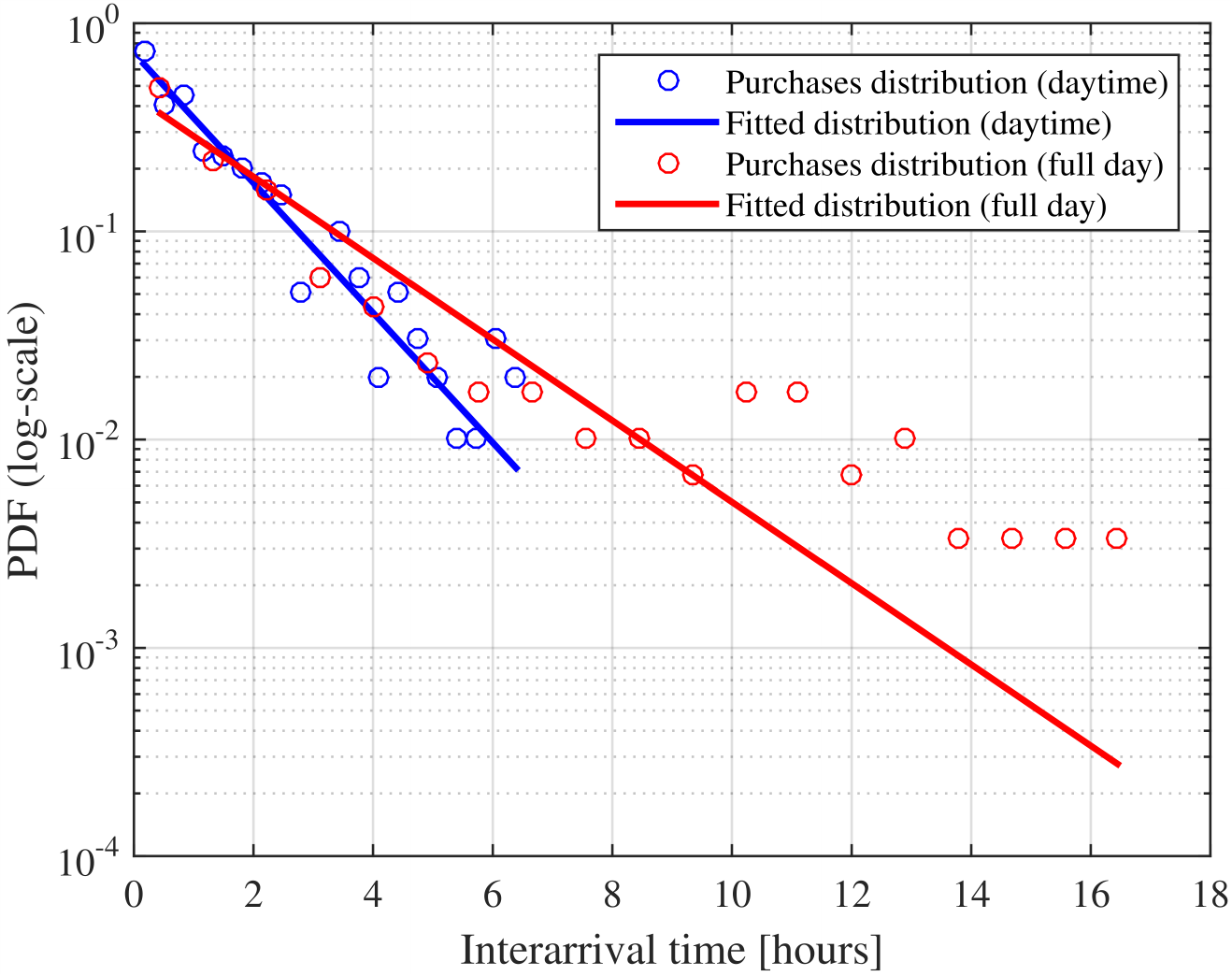}} \hfill
		\subfloat[]{\includegraphics[height=7.92cm]{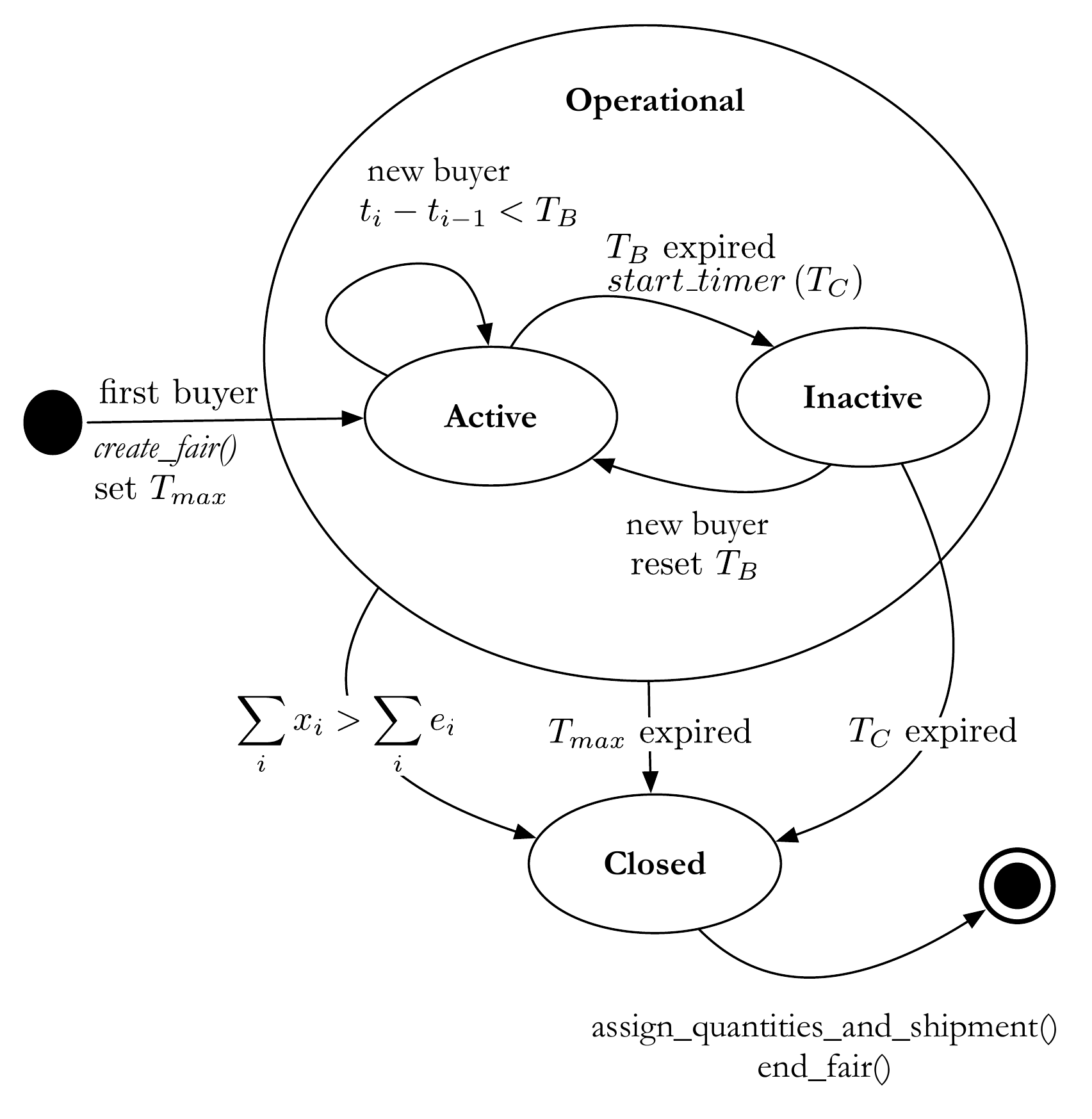}}
		\caption{System temporal evolutions:  inter-arrival time distribution for buyers (a); e-fair temporal model (b).}
		\label{f:fit_pdf}
	\end{figure*}
	The e-fair time analysis depends on buyers' arrival process and the e-fair time evolution.
	The first has been modeled using results of the Queuing Theory.
	Purchase events are considered as 'arrivals' of buyers in the system \cite{papoulis2002probability}.
	
	In traditional e-commerce, purchase inter-arrivals are exponentially distributed, as it is confirmed also by the literature and Kirivo's data. 
	However, even without fairs, inter-arrival processes are non-stationary.
	In facts, the arrival rate depends on the season (e.g. black Friday, Christmas holidays, periods of discounts) and on the time of day (arrivals are less frequent during nighttime). 
	Additionally, inter-arrivals are not truly independent because of the social diffusion of purchase information \cite{dynamics}.
	Focusing our study on a moving time window, we demonstrate that,
	in absence of fairs and during the daylight time, purchase inter-arrivals are almost exponentially distributed.
	Their probability density function (pdf) is given by $ f(x;\lambda) = \lambda e^{-\lambda x} u(x)$, where $\lambda$ is the arrival rate and $u(x)$ is the unit step function. 
	
	The empirical pdf of purchase inter-arrival during May 2015 is shown in Fig.~\ref{f:fit_pdf}.
	Inter-arrivals density over the 24 hours (in red dots) fits the exponential distribution except for inter-arrival times longer than 10 hours.
	This is due to purchase reduction during nighttime, that makes longer inter-arrivals more probable than provided by the exponential distribution.
	Reducing our analysis only to the daylight time (blue dots), we obtained a good fit with the exponential distribution.
	This null hypothesis has been verified at 5\% significance level, with the $\chi^2$-test, resulting in $p=0.8$ and $\chi^2=3.02$, with an excellent fit.
	Conversely, the number of arrivals in a given period of time is distributed as Poisson \cite{papoulis2002probability}.
	
\begin{figure}[tbp]
		\begin{center}
			
			\includegraphics[width=0.95\columnwidth, keepaspectratio = true]{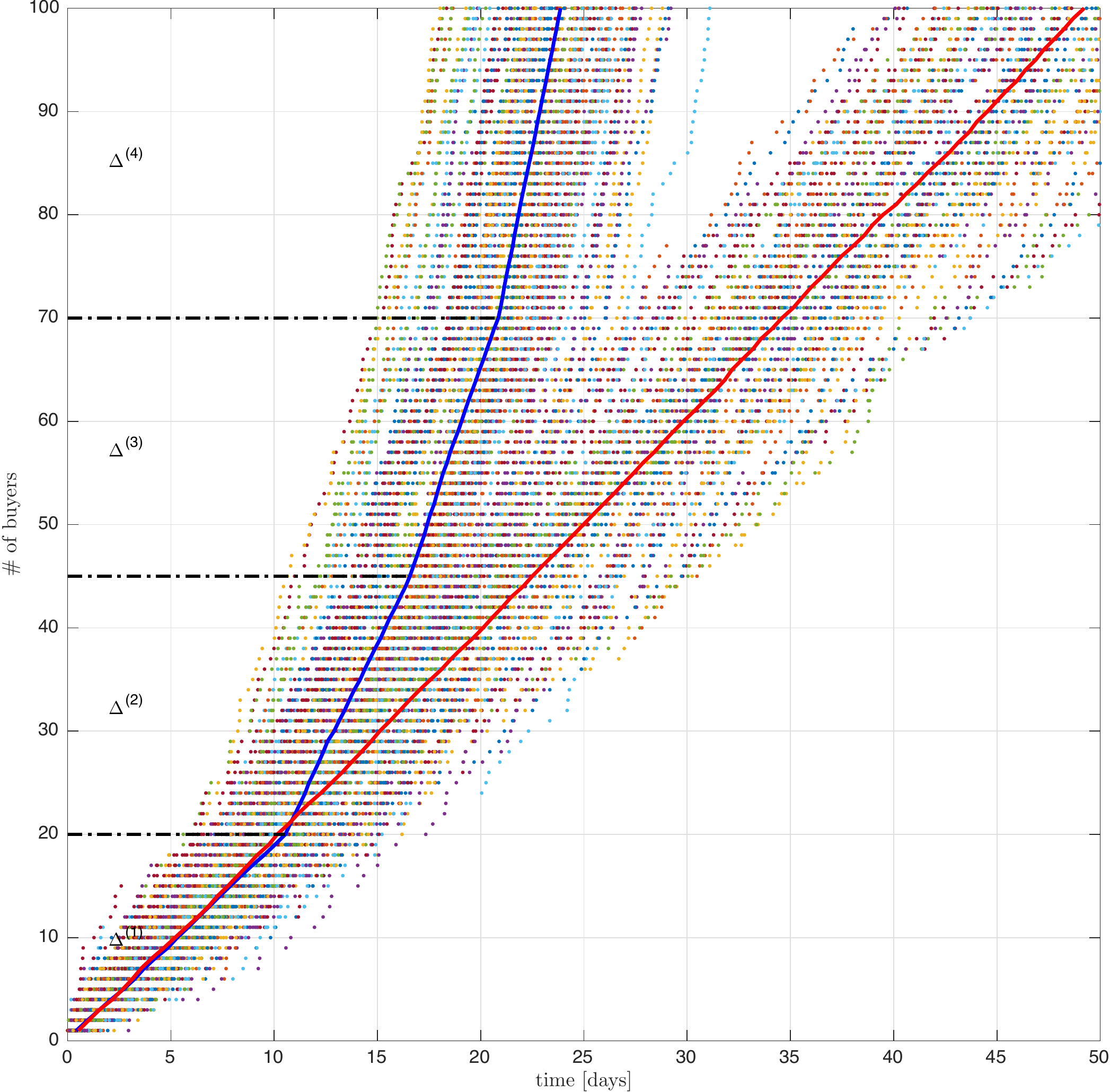}
			\caption{cumulated number of buyers' arrival: in red without fair and in blue with fair.}
			\label{f:earlangians}
		\end{center}
	\end{figure}

	Introducing e-fairs makes arrivals no more independent.
	In facts, buyers know the number of previous joins to the fair and can apply their own decision policy about their joining time.
	To model this phenomenon we consider that e-fairs with more buyers are more attractive than fairs with fewer buyers. 
	We modeled this behavior with the Markov chain depicted in Fig.~\ref{f:earlangians}a, where increasing birth rates ${\lambda_1 < \lambda_2 <\dots < \lambda_L}$ model such avalanche effect.
	It worths noting that $\lambda_l$ depends on the slope of the l-th piece within the price/quantity curve (see Fig.~\ref{f:piece-wise}).  
	The e-fair is then modeled as a birth process and buyers cannot leave the fair before its end.
	The status of the chain, $n$, indicates the number of buyers in the fair, and $p_n(t)$ denotes the probability to be in the status $n$ at time $t$.
	Applying the principle of flow conservation we derive the following differential equations:
	\begin{equation}
		\begin{cases} 
			p_0'(t)+    \lambda_1 p_0(t) =    0 \\ 
			p_n'(t)+    \lambda_1 p_n(t) =     \lambda_1 p_{n-1}(t) & 0 < n < n_1 \\ 
			p_{n_1}'(t)+\lambda_2 p_{n_1}(t)= \lambda_1 p_{n_1-1}(t) &  \\  
			p_n'(t)+    \lambda_2 p_n(t) =     \lambda_2 p_{n-1}(t) & n_1 < n < n_2  \\ 
			\vdots & \vdots \\    
			p_n'(t)+    \lambda_L p_n(t) =     \lambda_L p_{n-1}(t) & n_{L-1} < n < n_L\\
			p_{n_L}'(t)=\lambda_L p_{n_L-1}(t) & \\
		\end{cases}
	\end{equation}
	with initial conditions ${p_0(0)=1}$ and ${p_n(0)=0}$ for ${n>1}$, indicating that fairs are created without buyers.
	Furthermore, $n_1=\Delta_i^{(1)}$, $n_2=\Delta_i^{(1)}+\Delta_i^{(2)}$ and  $n_L=\sum_{l=1}^L \Delta_i^{(l)}$.
	
	These differential linear equations have to be resolved in cascade, being $p_n(t)$ necessary to find $p_{n+1}(t)$.
	The general solution provides the probability to have a given number of buyers at a specified time t and can be numerically found.
	However, as shown in Fig.~\ref{f:piece-wise}, for $x_i \leq n_1$ no purchase savings are obtained (shipment savings are still possible), while $x_i>n_1$ buyers obtain also price reductions. Therefore $n_1$ is a key value and the solution for the first $n_1+1$ equations is easy to find.
	In fact, it is a Poisson process with constant rate $\lambda_1$, therefore the solution is  \cite{papoulis2002probability}):
	\begin{equation}
		\label{e:erlangian}
		p_n(t)=\frac{e^{-\lambda_1 t} (\lambda_1 t)^n}{n!}
	\end{equation}
	In order to find the instant of time with the highest $p_n(t)$ probability, in our case for $n=n_1$, we maximize the expression in Eq.~\ref{e:erlangian} and obtain $t_1^*=\frac{n_1} {\lambda_1}$.
	
	\subsection{e-Fair time evolution}
	\label{s:architecture}    
	We modeled the e-fair evolution using the eXtended Finite State Machine (XFSM), reported in Fig.\ref{f:fit_pdf}B.
	e-Fairs can be active, inactive, or closed, plus the initial and final states, which are marked with special black circles.
	Transitions between states are depicted with arrows; they indicate the switch between old and new state. 
	These are characterized by \emph{events, conditions,} and \emph{actions}.
	Transitions are triggered by events (e.g. buyer arrival, products sold-out, timer expiration), then conditions are checked (e.g. inter-arrival time below/above a threshold, requested goods exceed the available ones) and actions (e.g. create\_fair(), set\_timer(), reset\_timer(), assign\_quantities(), ship(), optimize()).
	e-Fairs are \textit{Active} when buyers arrival rate is high enough, otherwise they become \textit{Inactive}.
	In both cases, the e-fair is \textit{Operational}, because it accepts new buyers' arrival.
	
	When new buyers arrive (except the first one), the optimization algorithm is run, which computes total prices, selects the involved seller(s) and quantities, the optimal shipping locations (see Fig. {f:fair-model}). 
	When the e-fair is operational, it is aggregating buyers. This macro-state 
	is buyer-driven and e-fairs can be operational for several days. 
	The fair switches to the closed state when the number of requested products exceeds their availability (cumulated over all sellers), the maximum fair duration exceeded, or a long period of inactivity occurred without new buyers' arrival. 

	\section{Experimental evaluation}
	In order to validate our model described in \S\ref{s:fair-based-model}, 
	we implemented a dedicated emulator using  MATLAB\textsuperscript{\textregistered} and run several 
	e-fair experiments.
	
	\subsection{Input data}
	We used several input data coming from the field, about POPs (location and 
	geographic distribution), purchases (time of arrival, purchased product, shipment address) and prices (price-quantity curves).
	
	A young Italian web marketplace named Kirivo provided anonymous real purchase data \cite{kirivo}.
	Price/quantity curves have been collected by asking them to selected sellers \cite{origini}.
	We evaluated the \emph{interest to buy} specific products using the number of clicks counted by trovaprezzi, an on-line price analysis system \cite{trovaprezzi}.
	Finally, the \emph{geographical distribution of POPs} is obtained from more than 1400 addresses of Fermo!Point POPs \cite{fermopoint}, whose geographical coordinates have been automatically extracted.
	
	We manipulated this input data for tackling the immaturity of the marketplace in use and for considering attraction effects introduced by e-fairs.
	Kirivo's volume of purchases increases over time and its growth is expected to expand with its reputation, therefore such data have valid trends but not significant absolute values.
	Because of this, we extracted time and geographical distributions of Kirivo's data and used them to create larger realistic datasets that better exploit e-fair aggregation. 
	As discussed in \S\ref{buyers-arrival} e-fairs impact on buyers' arrival time by introducing dependency, therefore we created arrival time accordingly to our Markov chain model.
	
	\begin{figure*}[!t]
		\centering
		\subfloat[]{\includegraphics[height=0.65\columnwidth, keepaspectratio = true]{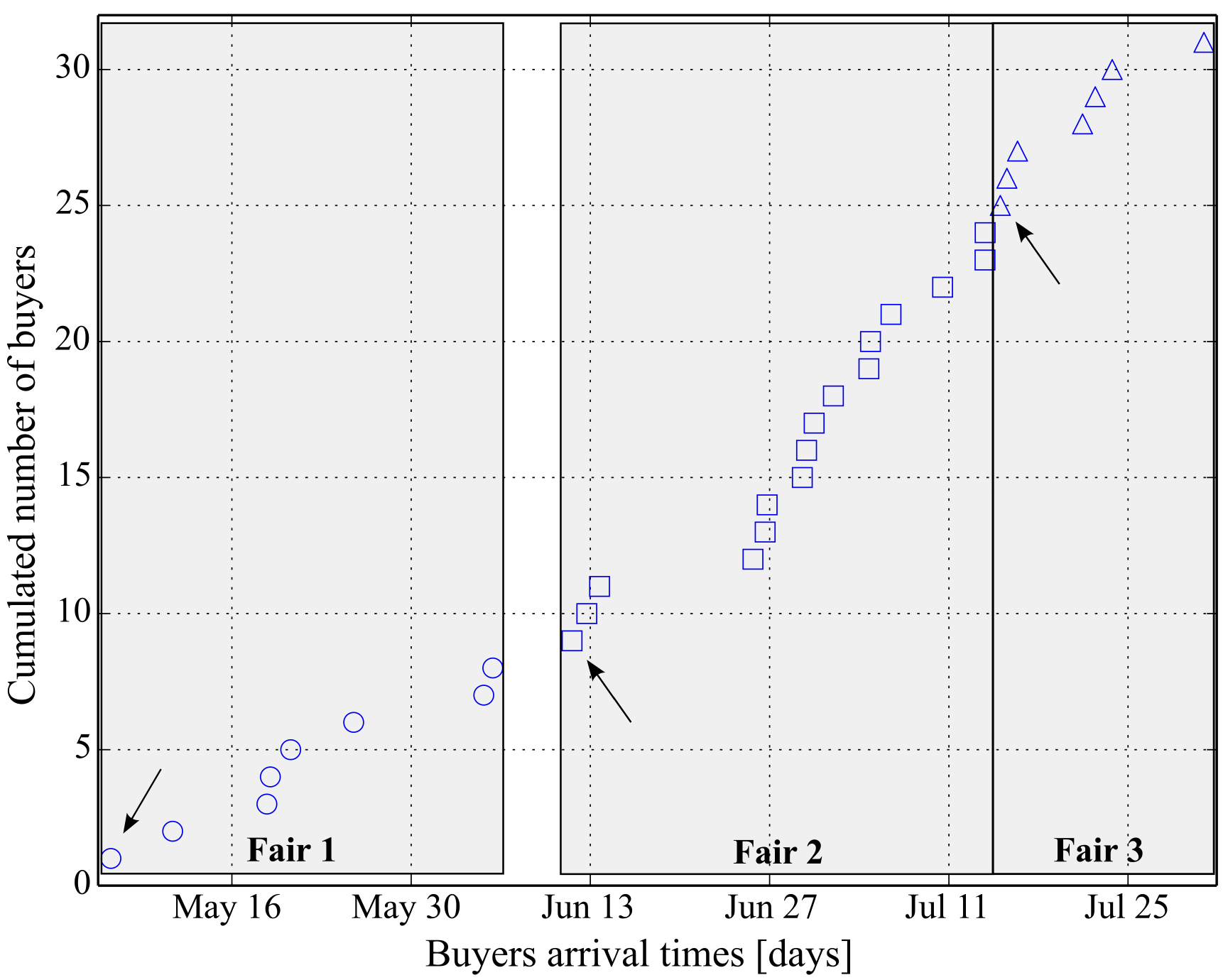}} \hfill
		\subfloat[]{\includegraphics[height=0.65\columnwidth,keepaspectratio = true]{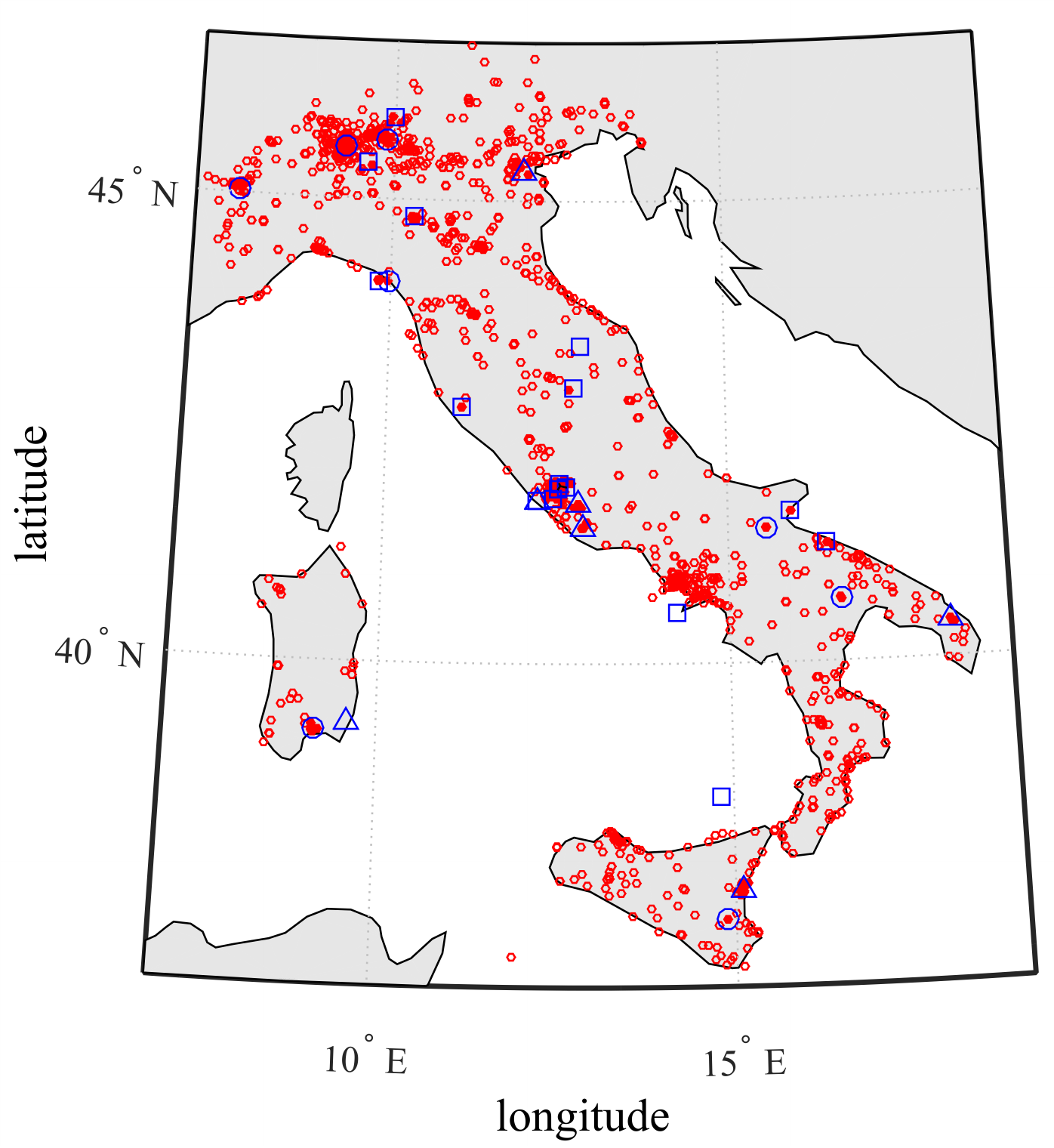}} \hfill
		\subfloat[]{\includegraphics[height=0.65\columnwidth, keepaspectratio = true]{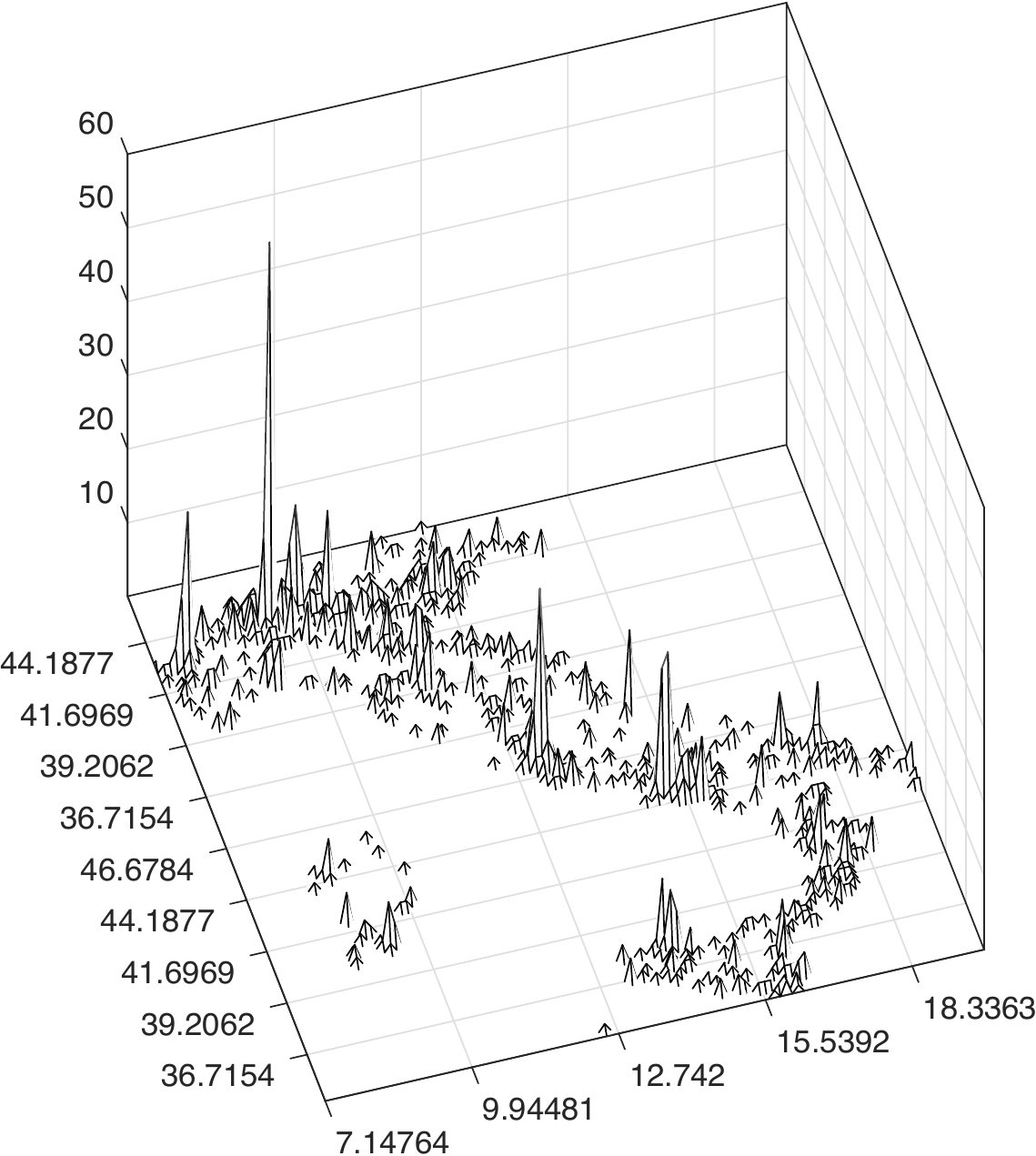}}
		\caption{Top selling product analysis: buyers' cumulative number of arrivals over time (a) and their geographical locations on the Italian territory (b); number of POPs distributed on a regular 2D grid in Italy (c).}
		\label{f:time-space-analysis}
	\end{figure*}
	We started validating fairs using the best-selling product because its highest arrival rate permits faster aggregations. 
	During a reference period of three months, the best-selling product on Kirivo was a smartphone of a popular brand.
	The cumulated number of purchases over time is shown in Fig.~\ref{f:time-space-analysis}a grouped within three gray boxes representing the three e-fairs. 
	The first e-fair is created when the first buyer arrives (see black arrows in the figure), then next buyers join the active fair until one terminating conditions occur.
	From the geographical point of view, in Fig.~\ref{f:time-space-analysis}b are shown locations of buyers and POPs, indicated with blue and red markers respectively.
	Furthermore, blue markers for buyers are distinguished with dots, squares and triangles depending on which of the three fairs they belong to Fig.~\ref{f:time-space-analysis}a,b.
	The geographical distribution of POPs, shown in Fig.~\ref{f:time-space-analysis}c,  appears to be closely related to population density and is used to optimize shipment consolidation and withdrawal costs.
	
	\subsection{Numerical results}
	In order to validate the feasibility and improvements introduced by fairs, we emulated several e-fairs and applied the optimization algorithm.
	Then we discuss timing and monetary results reported in graphical and tabular form respectively.
	
	Given the input data, the algorithm computes Haversine distances between buyers' and POPs' locations, 
	then it runs the pre-processing phase described in \S\ref{s:fair-based-model}, after that it separates the non-linear problem into an integer linear one, which is finally solved. 
	If a feasible solution is found, $x_{ij}$ and $y_{jk}$ are provided in addition to the total fair cost and the fraction to be paid by each buyer.
	Revenues for the e-fair management are also computed.
	
	We run 100 e-fairs and annotated buyers' arrival times. 
	Buyers' cumulative number is shown in Fig.~\ref{f:time-space-analysis}c, both with and without fair, indicated with dots.
	Averaging these results over time, we obtain the red diagram using legacy e-commerce (no e-fair), and the blue diagram using e-fair.
	In the first case, it is evident that buyers' arrival rate is constant in the period. 
	Conversely, the number of buyers has a faster growth in case of e-fair, and the cusps correspond to milestones in the number of buyers, where purchase savings change.
	In figure, the first cusp is at $t_1=10$ days, where $n_1=20$ buyers are expected.
	
	It worth noting that in this figure, slopes $\lambda_i$ depend on buyers and their demand, whereas $n_i$ are provided by the i-th seller because this value depends on his price/quantity curve.
	%
	
	Solutions for ${n \geq n_1}$ permit to find the probability to have a specific amount of savings on purchases because they depend on the number of buyers by expression \ref{eq:buyer_saving}.
	
	Used fair parameters are reported in tables \ref{tab:kirivo}a and \ref{tab:origini}a for Kirivo and Origini respectively.
	For the first set of experiments we used selling data from Kirivo marketplace and assumed the participation of 3 sellers, whose price / quantity simulated curves are described numerically in Table~\ref{tab:kirivo}b.
	Obtained results are shown in Table~\ref{tab:kirivo}c, where rows indicate fair output for any simulation.
	Rows contain the number of buyers that joined the fair, total costs and savings related to purchases and shipments (indicated with 'P' and 'S' respectively), as well as fair revenue. 
	The second experiment provides higher purchase savings than first and third because with 16 buyers, and therefore 16 demanded units of product, a better economy of scale is possible.
	Analogously, fair revenue is higher in second experiment (69.00\euro\ vs 17.25\euro\  and 17.25\euro).
	Furthermore, the second experiment provides also shipment and withdrawal savings.
	
	In the second set of experiments, we used the price / quantity curves provided by 3 sellers who are on Origini marketplace, and we have applied on simulated sales data.
	Each of these curves is referred to a different product sold, 
	we used one seller at a time to force no aggregation on the seller side.
	The number of bottles demanded by each buyer has been set to 8 units and in Table~\ref{tab:origini}c it results that with few buyers (i.e. 5, 10), the shipment savings is almost zero, because of the low probability of having buyers in proximity that can aggregate their withdrawals in one POP.
	Even savings from the purchase are always zero because the number of units required in both cases does not exceed $\Delta^{(1)}$.
	Having more than 15 buyers, significant price reductions are obtained for purchases and in shipments and savings increase up to 35.19\euro and 1.80\euro\ respectively.

\begin{center}
	\begin{table}[tbp]
		\begin{center}
			\caption{Fair parameters and numerical results using selling data from the field, for the Kirivo marketplace}
			\includegraphics[width=0.95\columnwidth, keepaspectratio = true]{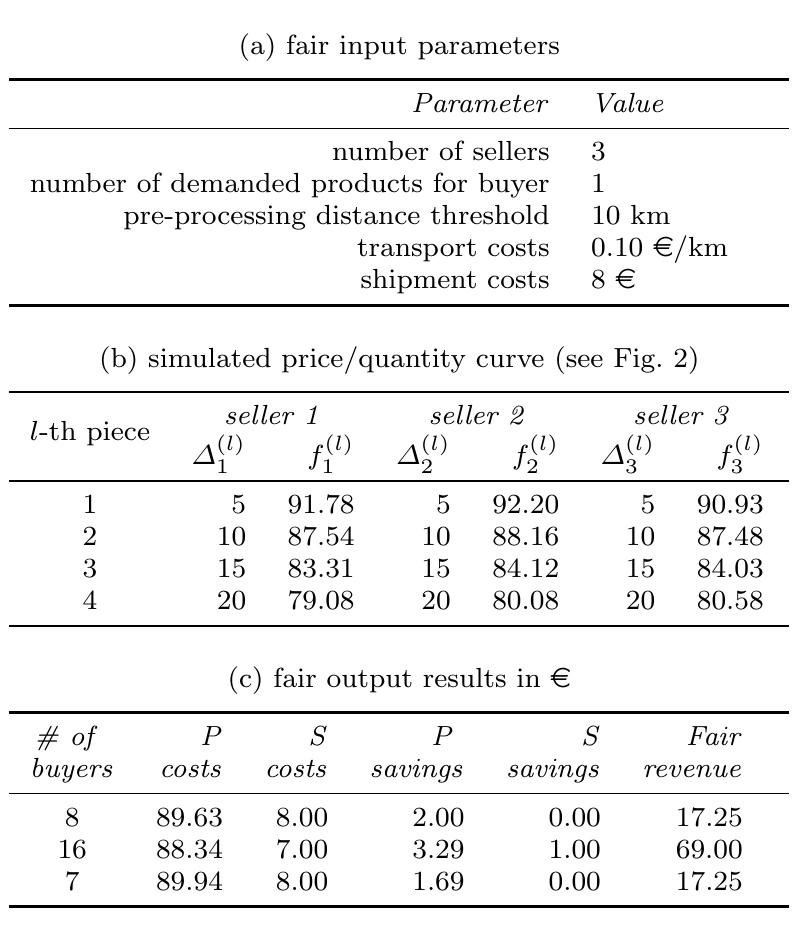}
			\label{tab:kirivo}
		\end{center}
	\end{table}	
\end{center}

\begin{center}	
	\begin{table}[tbp]
		\begin{center}
			\caption{Fair parameters and numerical results using price/quantity curves provided by Origini}
			\includegraphics[width=0.95\columnwidth, keepaspectratio = true]{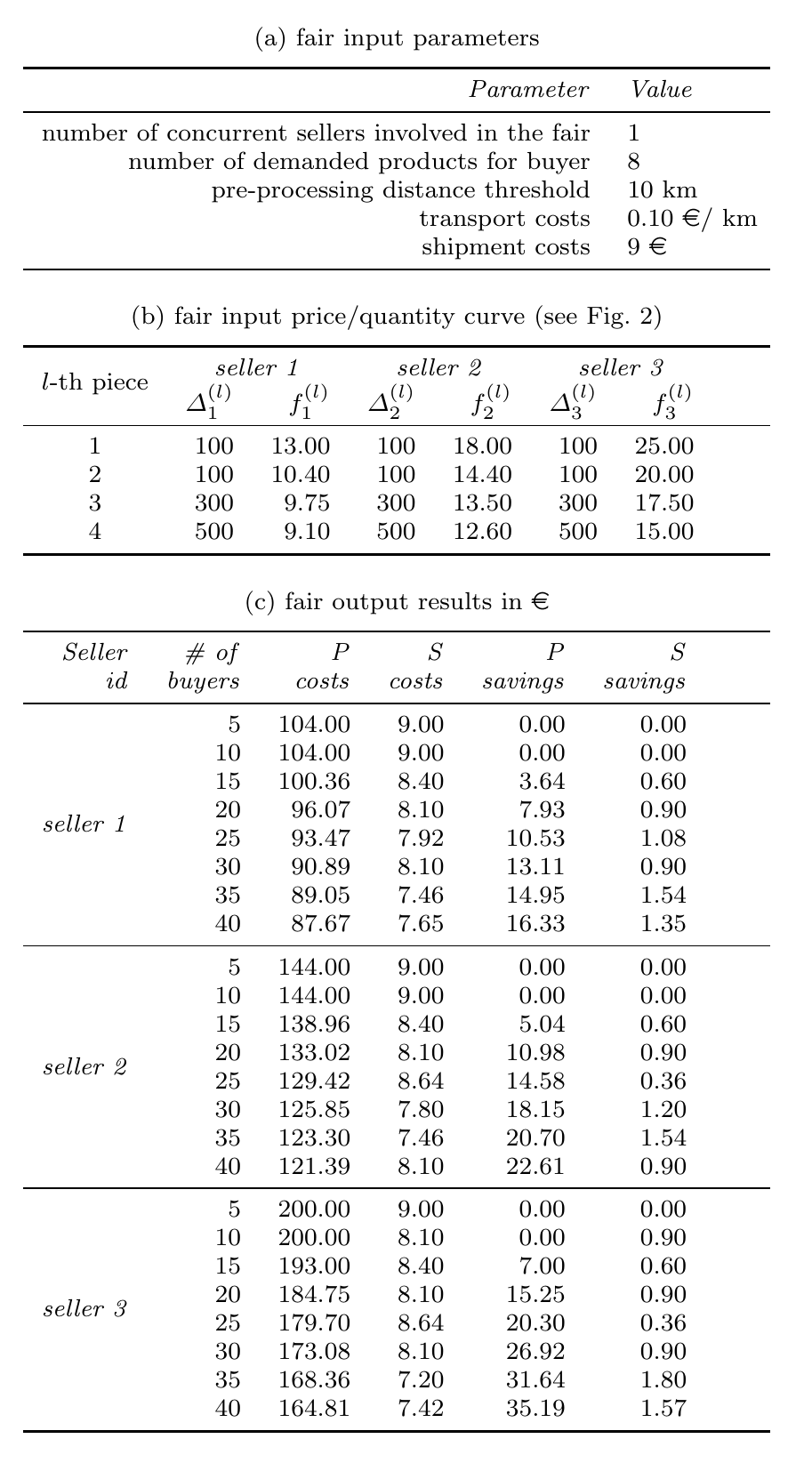}
			\label{tab:origini}
		\end{center}
	\end{table}	
\end{center}

	\section{Conclusion and future work}
	\label{s:conclusions}
	This paper presents a novel aggregation paradigm named e-fair, which applies to e-commerce and aggregates buyers' demands and sellers' offer while minimizing costs for purchases, shipments, and withdrawals. 
	
	e-Fairs go beyond group buying strategies and permit sellers and buyers to meet market requirements.
	We modeled e-fairs under several points of view (time, price, location) and by different perspectives (seller, buyer, e-fair manager).
	Unlike existing e-commerce models based on auctions, where prices increase due to competing buyers, our system leverages cooperative aggregation of buyers and permits prices downhill due to the economy of scale. 
	Additionally, e-fairs optimize shippings by opportunely consolidating them to a common point of delivery, which is in turns opportunely chosen to minimize withdrawal costs, afforded by buyers that retire their goods. 
	
	Every time new buyers join the e-fair, the system reruns the optimization algorithm, updates the optimal total cost and defines selected sellers provided quantities.
	Quantitative results demonstrate the efficacy of our approach and suggest its application in other scopes including services and energy.
	
	
	
	
	
	\section*{Acknowledgment}
	This work has been partially funded by 7Pixel S.r.l., which funds the Ph.D. scholarship held by Francesco Randazzo.
	
	\ifCLASSOPTIONcaptionsoff
	\newpage
	\fi
	
	\bibliographystyle{IEEEtran}
	\bibliography{bsbs-optimization}
	
	
	\begin{IEEEbiography}[{\includegraphics[width=1in,height=1.25in,clip,keepaspectratio]{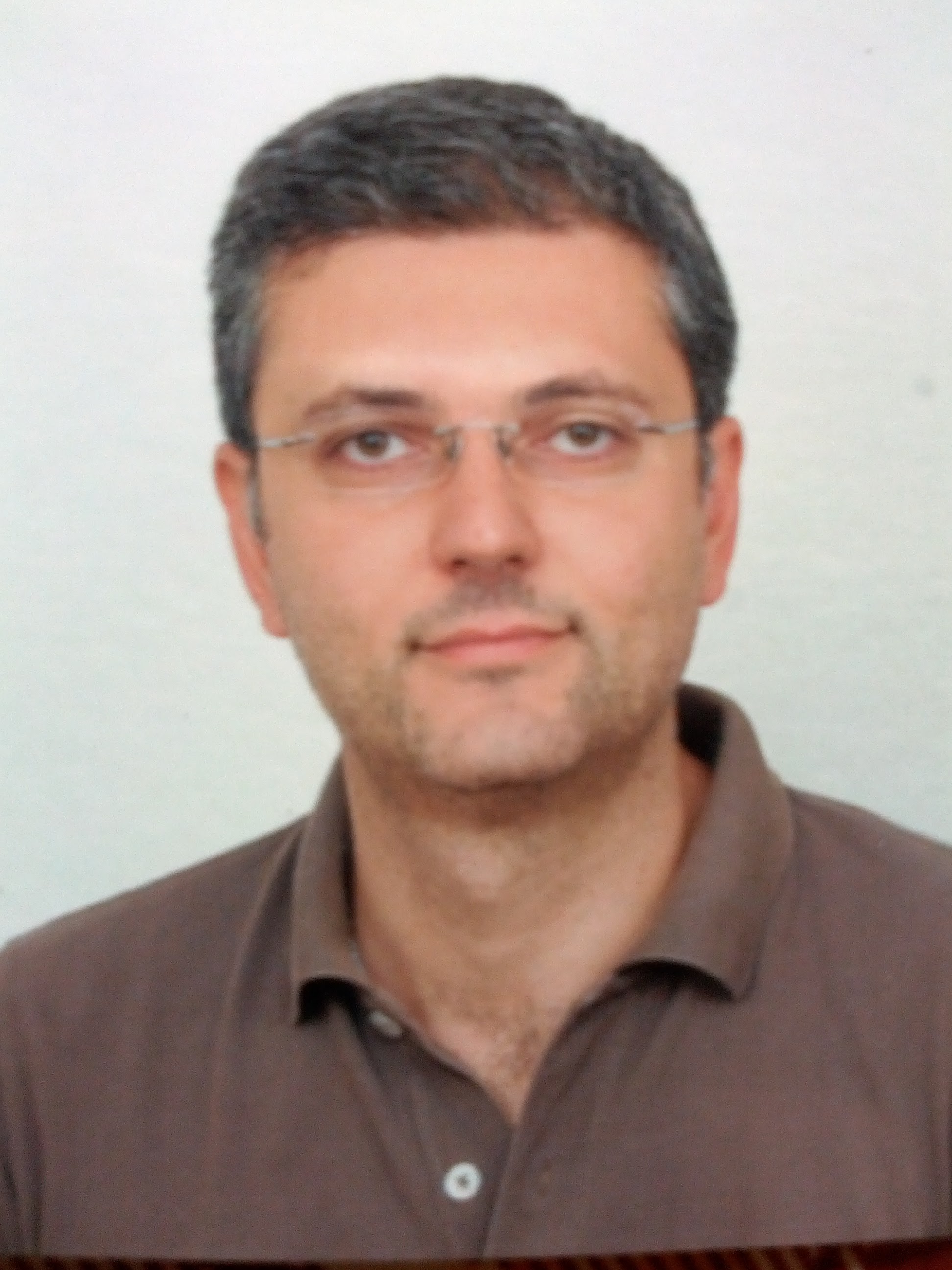}}]{Pierluigi Gallo} has been an Assistant Professor at the University of Palermo since November 2010.
		He graduated with distinction in Electronic Engineering in July 2002 and worked at CRES (Electronic Research Center in Sicily) until 2009. His work there was dedicated to QoS in IP core routers, IPv6 network mobility, and Wireless Networks. His research activity has focused on wireless networks at the MAC
		layer and 802.11 extensions, localization based on the time of arrival and cross-layer solutions. 
		P. Gallo has contributed to several national and European research projects: ITEA-POLLENS (2001-2003) on a middleware platform for a programmable router; IST ANEMONE (2006-2008) about IPv6 mobility; IST PANLAB II on the infrastructure implementation for federating testbeds; ICT FLAVIA (2010-2013) on Flexible Architecture for Virtualizable future wireless Internet Access; CREW (2013-2014) Cognitive Radio Experimental World; WiSHFUL(2015-) Wireless Software and Hardware platforms for Flexible and Unified radio and network control.
	\end{IEEEbiography}

	\begin{IEEEbiography}[{\includegraphics[width=1in,height=1.25in,clip,keepaspectratio]{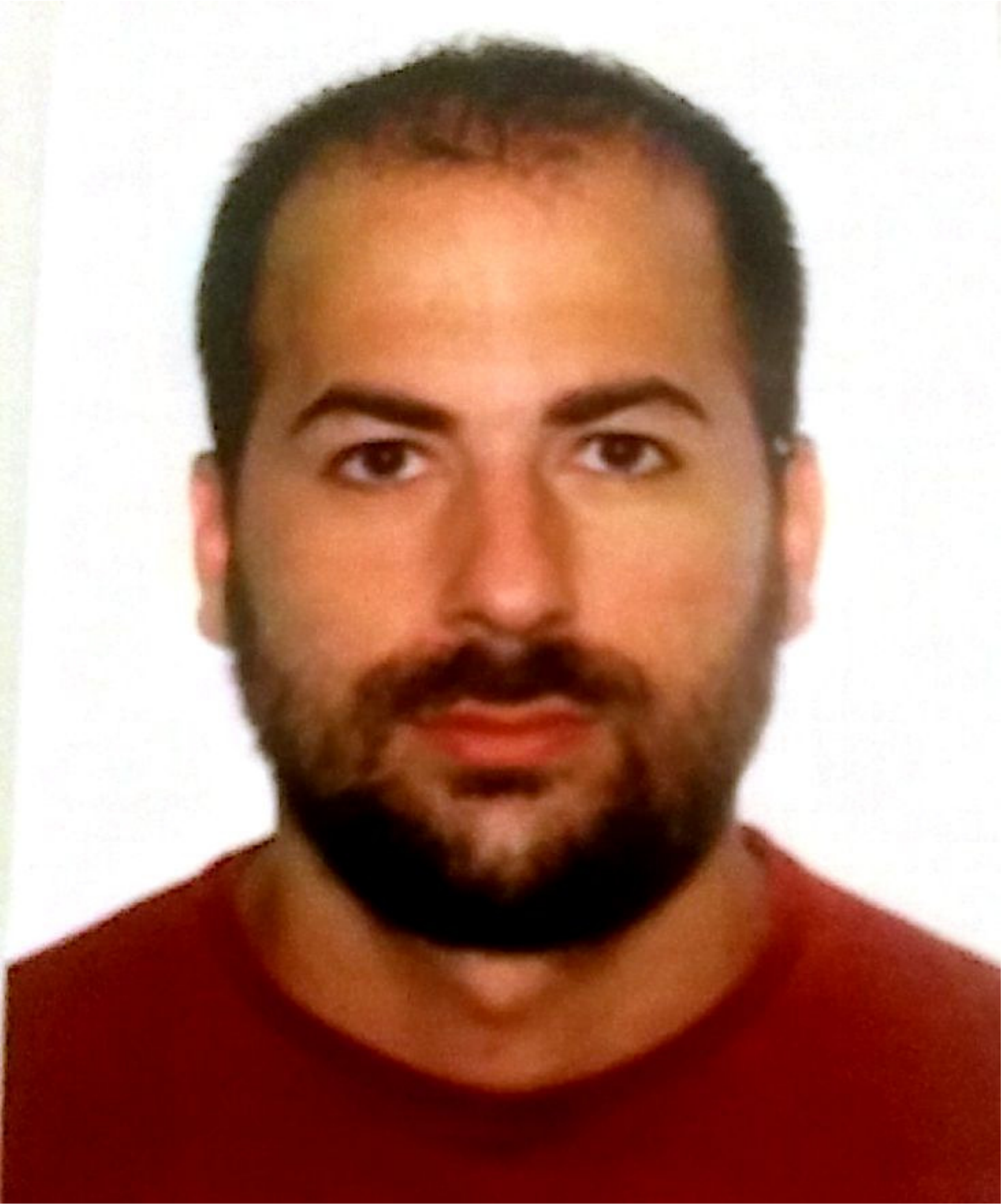}}]{Francesco Randazzo} has been a student at the University of Palermo. He graduated with first-class honours in Telecommunications Engineering in March 2015 and he is working at 7Pixel S.r.l since June 2015. His work is oriented to development of web applications for e-commerce.
	\end{IEEEbiography}
	\begin{IEEEbiography}[{\includegraphics[width=1in,height=1.25in,clip,keepaspectratio]{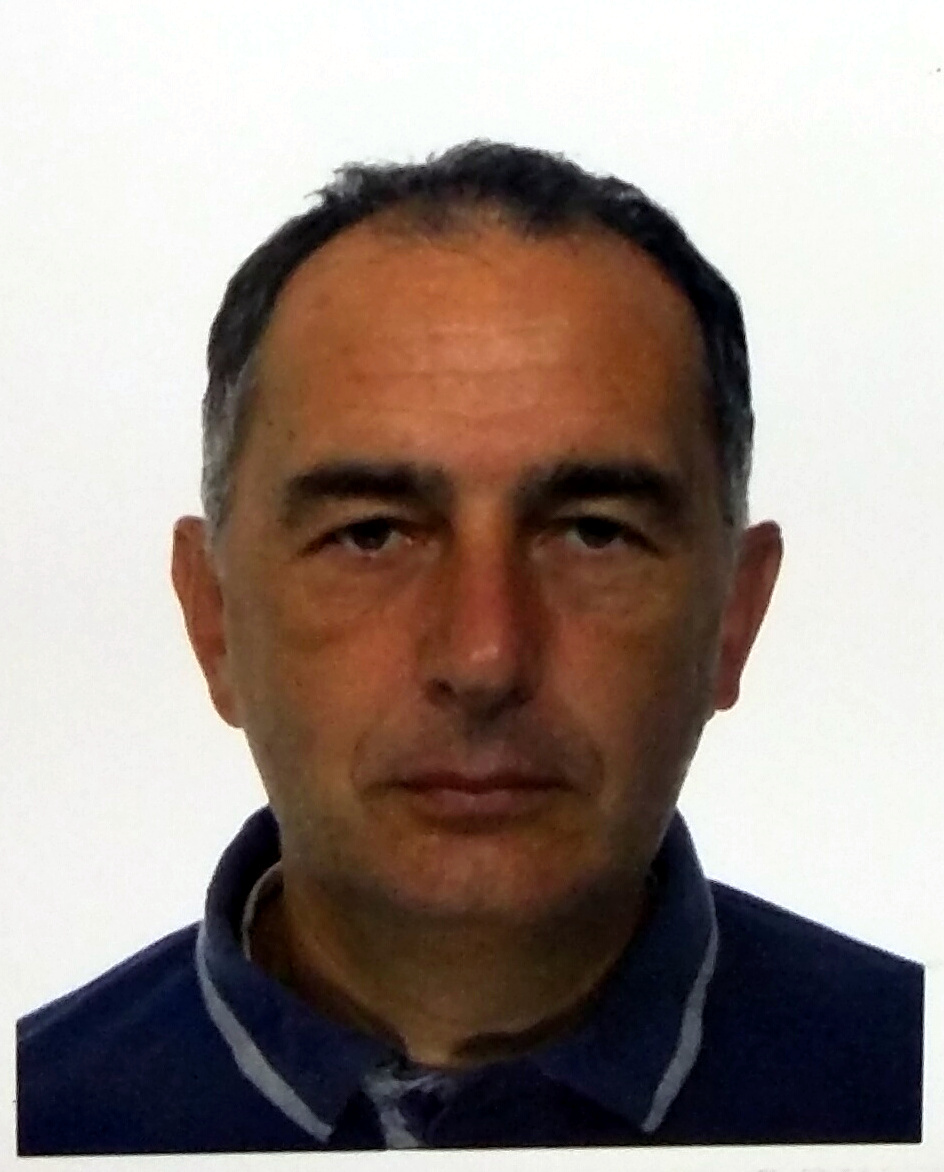}}]{Ignazio Gallo} 
		has been an Assistant Professor at the University of Insubria, Varese since 2003.
		He received his degree in Computer Science at the University of Milan, Italy, in 1998. 
		From 1998 to 2002 he worked at the National Research Council (CNR) in Milan for the Artificial Intelligence and Soft Computing laboratory.
		He was involved in the definition and development of neural models for classification and recognition of remote sensing image, and  neural models for decision support activities in engineering and environmental field.
		His research activity has focused on Computer Vision, Image Processing, Pattern Recognition, Neural Computing.
	\end{IEEEbiography}
	
	
	
	
\end{document}